\documentclass[12pt,english]{article}
\usepackage{mathptmx}
\usepackage[T1]{fontenc}
\usepackage{geometry}
\usepackage{comment}
\usepackage{color}
\usepackage{babel}
\usepackage{units}
\usepackage{mathtools}
\usepackage{amsmath}
\usepackage{amssymb}
\usepackage{stmaryrd}
\usepackage{setspace}
\usepackage{esint}
\usepackage{pgfplots}
\pgfplotsset{compat=1.12}
\usepackage{tikz}
\usepackage{physics}
\usepackage{array}
\usepgfplotslibrary{patchplots}
\usetikzlibrary{3d}
\usetikzlibrary{calc}
\usetikzlibrary{arrows,decorations.pathmorphing,backgrounds,positioning,fit,matrix}
\usepackage[toc,page]{appendix}
\usepackage{caption}
\usepackage{subcaption}

\usepackage[unicode=true,pdfusetitle,
 bookmarks=true,bookmarksnumbered=false,bookmarksopen=false,
 breaklinks=false,pdfborder={0 0 1},backref=false,colorlinks=true]
 {hyperref}
\hypersetup{
 linkcolor=blue,urlcolor={blue},filecolor={blue},citecolor={blue}}

\makeatletter

\newcommand{\lyxmathsym}[1]{\ifmmode\begingroup\def\b@ld{bold}
  \text{\ifx\math@version\b@ld\bfseries\fi#1}\endgroup\else#1\fi}

\usepackage{chngcntr}
\usepackage{mathtools}
\counterwithout{figure}{section}
\counterwithout{equation}{section}
\usepackage{times}
\usepackage{mwe}

\usepackage{times}
\usepackage[export]{adjustbox}

\counterwithout{figure}{section}
\counterwithout{equation}{section}
\usepackage{changepage}

\usepackage{babel}

\counterwithout{figure}{section}
\counterwithout{equation}{section}
\usepackage{times}
\usepackage{float}
\usepackage{graphicx}
\usepackage{bm}
\usepackage[numbers,sort&compress]{natbib} 
\usepackage{authblk}

\date{}

\renewcommand{\fnum@figure}{Figure \thefigure}

\makeatother

\begin{document}
\global\long\def\locs{x}%
\global\long\def\time{t}%
\global\long\def\ggt#1#2#3{#1_{#2#3}}%
\global\long\def\periodic#1{\hat{#1}}%
\global\long\def\derivative#1{#1{}_{,\locs}}%
\global\long\def\derivativet#1{#1{}_{,\time}}%
\global\long\def\effective#1{\left\langle #1\right\rangle }%
\global\long\def\ij#1#2#3{#1^{\left(#2,#3\right)}}%
\global\long\def\ijv#1#2{#1^{\left(#2\right)}}%
\global\long\def\effectiveprop#1{\tilde{#1}}%
\global\long\def\average#1{\bar{#1}}%
\global\long\def\dag#1{#1^{\dagger}}%
\global\long\def\trans#1{#1^{\mathsf{T}}}%
\global\long\def\reduced#1{#1_{s}}%
\global\long\def\reducedt#1{#1_{s,n}}%
\global\long\def\reducedtt#1{#1_{s,mn}}%
\global\long\def\three#1{\mathring{#1}}%
\global\long\def\dispv{\mathbf{u}}%
\global\long\def\momentumv{\mathbf{p}}%
\global\long\def\forcev{\mathbf{f}}%
\global\long\def\stresst{\boldsymbol{\sigma}}%
\global\long\def\elecdispv{\mathbf{D}}%
\global\long\def\elecfldv{\mathbf{E}}%
\global\long\def\strainsot{\boldsymbol{\eta}}%
\global\long\def\densityt{\boldsymbol{\rho}}%
\global\long\def\stiffnesst{\mathbf{C}}%
\global\long\def\B{\mathbf{B}}%
\global\long\def\A{\mathbf{A}}%
\global\long\def\couplingt{\mathbf{S}}%
\global\long\def\couplingS{S}%
\global\long\def\couplingW{W}%
\global\long\def\couplingelet{\mathbf{W}}%
\global\long\def\disp{u}%
\global\long\def\elecpot{\phi}%
\global\long\def\momentum{p}%
\global\long\def\force{f}%
\global\long\def\charge{q}%
\global\long\def\forces{\mathsf{f}}%
\global\long\def\stress{\sigma}%
\global\long\def\elecdisp{D}%
\global\long\def\elecfld{E}%
\global\long\def\strain{\epsilon}%
\global\long\def\strainso{\eta}%
\global\long\def\density{\rho}%
\global\long\def\stiffness{C}%
\global\long\def\Bb{B}%
\global\long\def\Aa{A}%
\global\long\def\coupling{\text{S}}%
\global\long\def\couplingele{\text{W}}%
\global\long\def\locationN{x}%
\global\long\def\location{\mathbf{x}}%
\global\long\def\reciprocal{G}%
\global\long\def\reciprocalvec{\mathbf{G}}%
\global\long\def\reciprocaltvec{\mathbf{G}'}%
\global\long\def\grad{\nabla}%
\global\long\def\div{\nabla\cdot}%
\global\long\def\curl{\nabla\times}%
\global\long\def\wavevec{\kappa}%
\global\long\def\angvel{\omega}%
\global\long\def\reciprocal{G}%
\global\long\def\reciprocalt{G'}%
\global\long\def\wavenum{\kappa}%
\global\long\def\kineticv{\mathsf{h}}%
\global\long\def\opcon{\mathsf{J}}%
\global\long\def\opgov{\mathsf{D}}%
\global\long\def\prop{\mathsf{L}}%
\global\long\def\Wmat{\mathsf{K}}%
\global\long\def\kinematicv{\mathsf{w}}%
\global\long\def\Q{\mathsf{Q}}%
\global\long\def\Qs{\breve{\mathsf{Q}}}%
\global\long\def\R{\mathsf{R}}%
\global\long\def\G{\left(G\right)}%
\global\long\def\Gt{\left(G'\right)}%
\global\long\def\Gm{\left(-G\right)}%
\global\long\def\Gtm{\left(-G'\right)}%
\global\long\def\GGt{\left(G-G'\right)}%
\global\long\def\strainsov{\mathsf{m}}%
\global\long\def\kG{\left(\wavenum+G\right)}%
\global\long\def\kGt{\left(\wavenum+G'\right)}%
\global\long\def\sumgt{\underset{\reciprocalt\ne0}{\sum}}%
\global\long\def\unitv{\mathbf{d}}%
\global\long\def\unit{l_T}%
\global\long\def\sumGt{\underset{\reciprocalt}{\sum}}%
\global\long\def\sumG{\underset{\reciprocal}{\sum}}%
\global\long\def\sumGv{\underset{\reciprocalvec}{\sum}}%
\global\long\def\sumGtv{\underset{\reciprocaltvec}{\sum}}%
\global\long\def\properties{\xi}%
\global\long\def\hfield{\alpha}%
\global\long\def\wfield{\beta}%
\global\long\def\fieldv{\mathbf{h}}%
\global\long\def\sources{s}%
\global\long\def\sourcescon{s^{0}}%
\global\long\def\sumGtG{\underset{G,\reciprocalt}{\sum}}%
\global\long\def\sumGtGv{\underset{\reciprocalvec,\reciprocaltvec}{\sum}}%
\global\long\def\diagG{\mathsf{Y}}%
\global\long\def\diagGt{\mathsf{Y}^{'}}%
\global\long\def\diagw{\mathsf{Y}^{\angvel}}%
\global\long\def\DL{\mathsf{T}}%

\global\long\def\g#1{#1_{\reciprocal}}%
\global\long\def\zero#1{#1_{0}}%
\global\long\def\gt#1{#1_{\reciprocalt}}%
\global\long\def\coeffg#1{\check{#1}_{\reciprocal}}%
\global\long\def\coeffgv#1{\check{#1}_{\reciprocalvec}}%
\global\long\def\coeffgt#1{\check{#1}_{\reciprocalt}}%
\global\long\def\ijggt#1#2#3{#1_{\reciprocal\reciprocalt}^{\left(#2,#3\right)}}%
\global\long\def\coeffggt#1{\check{#1}_{\reciprocal\reciprocalt}}%
\global\long\def\assembly#1{#1_{\mathrm{A}}}%
\global\long\def\numwaves{N}%
\global\long\def\ijvg#1#2{#1_{\reciprocal}^{\left(#2\right)}}%
\global\long\def\ijvgt#1#2{#1_{\reciprocalt}^{\left(#2\right)}}%
\global\long\def\coeffz#1{\check{#1}_{0}}%

\global\long\def\x{x_{1}}%
\global\long\def\y{x_{2}}%
\global\long\def\z{x_{3}}%
\global\long\def\xderivative#1{#1{}_{,\mathrm{1}}}%
\global\long\def\yderivative#1{#1{}_{,2}}%
\global\long\def\xst{\stress^{13}}%
\global\long\def\yst{\stress^{23}}%
\global\long\def\xD{\elecdisp^{1}}%
\global\long\def\yD{\elecdisp^{2}}%
\global\long\def\xstr{\strainso^{1}}%
\global\long\def\ystr{\strainso^{2}}%
\global\long\def\Ccon{c}%
\global\long\def\Bcon{b}%
\global\long\def\Acon{a}%
\global\long\def\xcoest{\coeffg{\stress}^{13}}%
\global\long\def\ycoest{\coeffg{\stress}^{23}}%
\global\long\def\xcoeD{\coeffg{\elecdisp}^{1}}%
\global\long\def\ycoeD{\coeffg{\elecdisp}^{2}}%
\global\long\def\xperst{\periodic{\stress}^{13}}%
\global\long\def\yperst{\periodic{\stress}^{23}}%
\global\long\def\xperD{\periodic{\elecdisp}^{1}}%
\global\long\def\yperD{\periodic{\elecdisp}^{2}}%
\global\long\def\kone{k_{1}}%
\global\long\def\ktwo{k_{2}}%
\global\long\def\xcoestr{\coeffgt{\strainso}^{1}}%
\global\long\def\ycoestr{\coeffgt{\strainso}^{2}}%
\global\long\def\xperstr{\periodic{\strainso}^{1}}%
\global\long\def\yperstr{\periodic{\strainso}^{2}}%
\global\long\def\Cm{\mathsf{\stiffness}}%
\global\long\def\Bm{\mathsf{\Bb}}%
\global\long\def\Am{\mathsf{\Aa}}%
\global\long\def\coupm{\mathsf{\coupling}}%
\global\long\def\coupelem{\mathsf{\couplingele}}%
\global\long\def\coeffgv#1{\check{#1}_{\reciprocalvec}}%
\global\long\def\coeffgtv#1{\check{#1}_{\reciprocaltvec}}%
\global\long\def\coeffggtv#1{\check{#1}_{\reciprocalvec\reciprocaltvec}}%
\global\long\def\xcoestv{\coeffgv{\stress}^{13}}%
\global\long\def\ycoestv{\coeffgv{\stress}^{23}}%
\global\long\def\xcoeDv{\coeffgv{\elecdisp}^{1}}%
\global\long\def\ycoeDv{\coeffgv{\elecdisp}^{2}}%
\global\long\def\xcoestrv{\coeffgtv{\strainso}^{1}}%
\global\long\def\ycoestrv{\coeffgtv{\strainso}^{2}}%
\global\long\def\Gone{\reciprocal_{1}}%
\global\long\def\Gtwo{\reciprocal_{2}}%
\global\long\def\Gtone{\reciprocalt_{1}}%
\global\long\def\Gttwo{\reciprocalt_{2}}%
\global\long\def\xavst{\average{\stress}^{13}}%
\global\long\def\yavst{\average{\stress}^{23}}%
\global\long\def\xavD{\average{\elecdisp}^{1}}%
\global\long\def\yavD{\average{\elecdisp}^{2}}%
\global\long\def\xzerost{\coeffz{\stress}^{13}}%
\global\long\def\yzerost{\coeffz{\stress}^{23}}%
\global\long\def\xzeroD{\coeffz{\elecdisp}^{1}}%
\global\long\def\yzeroD{\coeffz{\elecdisp}^{2}}%
\global\long\def\sfzero{\mathsf{0}}%
\global\long\def\BbT{B_{G'}^{T}}%

\title{Maximizing the electromomentum coupling in piezoelectric laminates}
\author[1]{Majd Kosta}
\author[2]{Alan Muhafara}
\author[2]{Rene Pernas-Salómon\footnote{Present address: Universidad Carlos III de Madrid,Avenida de la Universidad 30,28911, Leganés, Spain.}}
\author[2]{Gal Shmuel}
\author[1]{Oded Amir}
\affil[1]{\small{Faculty of Civil Engineering, Technion--Israel Institute of Technology,
Haifa 32000, Israel}}
\affil[2]{\small{Faculty of Mechanical Engineering, Technion--Israel Institute of
Technology, Haifa 32000, Israel}}

\maketitle


\begin{abstract}

Asymmetric piezoelectric composites exhibit coupling between their macroscopic linear momentum and electric field, a coupling that does not appear at the microscopic scale. This electromomentum coupling constitutes an additional knob to tailor the dynamic response of the medium, analogously to the Willis coupling in elastic composites. Here, we employ topology- and free material optimization approaches to maximize the electromomentum coupling of periodic piezoelectric laminates in the low frequency, long-wavelength limit. We find that the coupling can be enhanced by orders of magnitude, depending on the degrees of freedom in the optimization process. The optimal compositions that we find provide guidelines for the design of metamaterials with maximum electromomentum coupling, paving the way for their integration in wave control applications.

\end{abstract}
\section{Introduction}
Metamaterials are artificial composites whose effective behavior is fundamentally different from the behavior of their constituents \cite{Kadi2019nrp,Christensen2015MRCComunications,craster2012acoustic,deymier2013acoustic,Ma2016SA,Simovski2009bc,Smith2006JOSA,lustig2019,Sridhar2018JMPS,Wegener2013}. Of particular relevance to this work are Willis metamaterials \cite{Willis1985IJSS,Willis1981WM,willis1981variational,Meng2018prsa,Milton07,Muhlestein20160Prsa2,nassar2015willis,Norrisrspa2011PRSA,quan2018prl,Shuvalov2011prsa,Sieck2017prb,Srivastava2015prsa,Ponge2017EML,Torrent2015PRB,Chen2020nc,Lau2019,Melnikov2019nc,Popa2018nc,Muhlestein2017nc,Liu2019prx,Merkel2018prb}, whose linear momentum and strain is coupled, and so are their  stress and velocity: these couplings do not appear in homogeneous media. They are captured by the so-called Willis tensor that enters the effective constitutive relations of the composite, reflecting a designable degree of freedom to tailor its dynamic response. The Willis tensor is nonlocal in space and time; its (spatially) local version was introduced by \citet{milton06cloaking}, and a model that demonstrates such behavior was developed later by    \citet{Milton2007njp}.

Recently, \citet{PernasSalomon2019JMPS} have generalized the Willis couplings to piezoelectric composites, discovering that their linear momentum and electric displacement field can be macroscopically coupled with the electric field and velocity, respectively.
This effect is captured by a second-order tensor—the \textit{electromomentum} tensor—that  enters the effective constitutive relations of the piezoelectric composites.  Analogously to the Willis tensor, the electromomentum tensor not only constitutes an additional knob to tailor the dynamic response of the medium, but is also necessary for obtaining a physical constitutive description \cite{pernas2020fundamental,rps20201wm,muhafra2021}. 

The objective of this work is to maximize the electromomentum effect in periodic piezoelectric laminates driven by axial load sources. The poling direction of the piezoelectric constituents is also in the lamination direction, such that the problem is one-dimensional, and the electromomentum tensor becomes a scalar material property. We further restrict attention to the low frequency, long-wavelength limit, where the electromomentum coefficient depends linearly on the frequency, and is local in space. The motivation for identifying laminates with large electromomentum coefficient is twofold. First, it makes more pronounced the phenomena that the electromomentum effect generate, such as the change in the phase velocity and directional phase angle of elastic waves \cite{rps20201wm}. Second, the optimal compositions are expected to elucidate the dependency of the electromomentum coupling on the mechanical and geometrical properties of the constituents. Very recently, a preliminary work towards this end was done by  \citet{zhang2022rational}. There, the authors  studied 8-layer unit cells made of five real materials, and examined how the electromomentum coefficient changes as function of  the different material combinations and layer thickness. Here, we take a different route,  using optimization methods. 
Specifically, our approach is based on topology- and free material optimization (FMO) methods: topology optimization aims to determine the optimal material distribution for a designated objective function in a given design space \cite{bendsoe_topology_2004};  FMO \cite{bendsoe1994analytical, zowe1997free, kovcvara2008free} is an extension of topology optimization, in which each material property can vary independently of the other properties. Since the design space for FMO is free from the constraint of real materials, it provides a kind of a theoretical upper bound for the objective function. 

The objective function, i.e., the electromomentum coefficient, is formulated as the end result of a homogenization process. In this process, the heterogeneous constitutive properties of the composite are replaced by fictitious homogeneous constitutive properties, relating the macroscopic kinematic variables to the macroscopic kinetic variables. Further details regarding the homogenization scheme are given in Section \ref{chap:PWE}. Accordingly, the optimization methods that we apply are in fact inverse homogenization methods \cite{sigmund1994materials} for the constitutive property. 

Our analysis begins in Section \ref{sec:DMO} with discrete material optimization, or DMO \cite{stegmann2005discrete, niu2010discrete}, where the design space is a predefined set of real materials. For simplicity, here and in the subsequent optimization problems, we restrict attention to periodic cells that are made of three layers. Since this optimization is based on a small number of discrete variables, we use a genetic algorithm to solve the optimization problem. We find that the electromomentum coefficient of the optimal laminate is two orders of magnitude larger than the laminate that was first considered by \citet{PernasSalomon2019JMPS}, which was chosen arbitrarily. Henceforth, we refer to the latter laminate as the reference laminate. We observe that the optimization yields a choice of materials that tends to maximize the contrast between their electromechanical properties, in a unit cell that comprises one thin layer in-between two thick layers. 

In Section \ref{sec:GradientBased} we consider a design space with continuous variables, whose values are bounded in-between the extreme values of the materials considered earlier. We solve different optimization problems using gradient-based algorithms that are implemented together with a sensitivity analysis of the objective function. We first consider optimization problems whose design variables are pairs that consist the mass density and one of the electromechanical properties, i.e., the dielectric-, piezoelectric- and stiffness coefficients, while all the rest of the properties are set to the properties of the reference laminate. We find that each one of the optimal pairs enlarge the electromomentum coefficient by another order of magnitude, in comparison with the optimal laminate that is obtained from the DMO. We then proceed and increase the number of design variables, and find that when all material properties are designable, the electromomentum coefficient is enlarged by another order of magnitude. We conclude this paper in Section \ref{sec:sum}, with a summary of our main results.



\section{The homogenization process}
\label{chap:PWE}


\begin{figure}[t!] 
\centering
\usetikzlibrary{quotes,arrows.meta}
\tikzset{
  annotated cuboid/.pic={
    \tikzset{%
      every edge quotes/.append style={midway, auto},
      /cuboid/.cd,
      #1
    }
    \draw [every edge/.append style={pic actions, densely dashed, opacity=.5}, pic actions]
    (0,0,0) coordinate (o) -- ++(-\cubescale*\cubex,0,0) coordinate (a) -- ++(0,-\cubescale*\cubey,0) coordinate (b) edge coordinate [pos=1] (g) ++(0,0,-\cubescale*\cubez)  -- ++(\cubescale*\cubex,0,0) coordinate (c) -- cycle
    (o) -- ++(0,0,-\cubescale*\cubez) coordinate (d) -- ++(0,-\cubescale*\cubey,0) coordinate (e) edge (g) -- (c) -- cycle
    (o) -- (a) -- ++(0,0,-\cubescale*\cubez) coordinate (f) edge (g) -- (d) -- cycle;
  },
  /cuboid/.search also={/tikz},
  /cuboid/.cd,
  width/.store in=\cubex,
  height/.store in=\cubey,
  depth/.store in=\cubez,
  units/.store in=\cubeunits,
  scale/.store in=\cubescale,
  width=10,
  height=10,
  depth=10,
  units=cm,
  scale=.1,
}

\begin{tikzpicture}
  \pic [fill=blue, text=blue, draw=blue] at (0,0,0) {annotated cuboid={width=5, height=5, depth=15}};
  \pic [fill=gray, text=gray, draw=gray] at (0,0,1) {annotated cuboid={width=5, height=5, depth=10}};
  \pic [fill=orange, text=orange, draw=orange] at (0,0,2) {annotated cuboid={width=5, height=5, depth=10}};
  
  \pic [fill=blue, text=blue, draw=blue] at (0,0,3) {annotated cuboid={width=5, height=5, depth=15}};
  \pic [fill=gray, text=gray, draw=gray] at (0,0,4) {annotated cuboid={width=5, height=5, depth=10}};
  \pic [fill=orange, text=orange, draw=orange] at (0,0,5) {annotated cuboid={width=5, height=5, depth=10}};
  
  \pic [fill=blue, text=blue, draw=blue] at (0,0,6) {annotated cuboid={width=5, height=5, depth=15}};
  \pic [fill=gray, text=gray, draw=gray] at (0,0,7) {annotated cuboid={width=5, height=5, depth=10}};
  \pic [fill=orange, text=orange, draw=orange] at (0,0,8) {annotated cuboid={width=5, height=5, depth=10}};
  
  \pic [fill=blue, text=blue, draw=blue] at (0,0,9) {annotated cuboid={width=5, height=5, depth=15}};
    
  \draw[latex'-latex'] (0.2,-0.5,0) -- (0.2,-0.5,1)
  node[pos=0.25,below]{$l_3$};
  
  \draw[latex'-latex'] (0.2,-0.5,1) -- (0.2,-0.5,2)
  node[pos=0.25,below]{$l_2$};
  
  \draw[latex'-latex'] (0.2,-0.5,2) -- (0.2,-0.5,3)
  node[pos=0.25,below]{$l_1$};

  \draw[latex'-latex'] (0.2,-0.5,6) -- (0.2,-0.5,9)
  node[pos=0.45,below]{$l_T$};
  
  \draw[|-stealth] (1,-0.5,9) -- (1,-0.5,7.5)
  node[pos=0.45,below]{$x$};

  \draw [thick,dashed] (0,0,9) -- (0,0,6);
  \draw [thick,dashed] (-0.5,0,9) -- (-0.5,0,6);
  
  \draw [thick,dashed] (0,-0.5,9) -- (0,-0.5,6);
  
  \draw [thick,dashed] (0,0,9) -- (0,-0.5,9);
  \draw [thick,dashed] (0,0,6) -- (0,-0.5,6);
  
  \draw [thick,dashed] (-0.5,0,6) -- (0,-0,6);
  \draw [thick,dashed] (-0.5,0,9) -- (0,-0,9);
  \draw [thick,dashed] (-0.5,-0.5,9) -- (-0.5,0,9);
  
  \draw [thick,dashed] (-0.5,-0.5,9) -- (0,-0.5,9);
\end{tikzpicture}
\caption{Part of an infinite medium made of a periodic cell with three layers. The length of the first, second and third layer is denoted by $l_1$, $l_2$ and $l_3$, respectively. The elementary unit cell is boxed in dashed lines, where $l_{T} = \sum_{m=1}^{3} l_{m}$. }
\label{fig:ThreeLayersExample}
\end{figure}
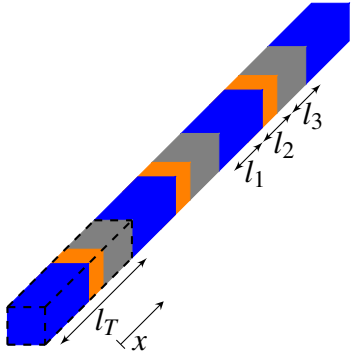
The homogenization process that we use to extract the electromomentum coefficient is founded on two principles. (\emph{i}) The macroscopic fields are defined by the product of the Bloch envelope of the microscopic fields and the volume average of their periodic part.  Defined in this way, the effective fields satisfy identically macroscopic governing equations that are of the same form as the microscopic equations, see Refs.\ \cite{Willis2011PRSA,PernasSalomon2019JMPS,alu2011first}. (\emph{ii}) We account for several driving sources in the formulation \cite{Fietz2009,FIETZ2010pysicaaB,alu2011prb,Willis2011PRSA,PernasSalomon2019JMPS}; collectively, these sources allow us to obtain a \emph{unique} set of effective properties \cite{WILLIS2012MRC,nassar2015willis,pernassalomn2020prapplied,Milton2020IV}, which satisfies necessary  physical laws \cite{Alu2011-PhysRevB,Muhlestein20160Prsa2,pernassalomn2020prapplied,muhafra2021,pernas2021electromomentum}. 

Our averaging process uses the plane wave
expansion method \cite{Kushwaha1993,SigalasEconomou96,Norrisrspa2011PRSA,Ponge2017EML,pernassalomon2018jmps}, and the resultant scheme is essentially a one-dimensional reduction of the scheme that was developed by \citet{muhafra2021}. In order to make this manuscript self-contained and introduce the quantities that are used in the optimization process, the complete one-dimensional scheme is provided next.   

Consider a periodic repetition of three piezoelectric layers, the poling direction of which is in the lamination direction, say $x$. The laminate is driven
by a body force density $f$, axial inelastic strain $\eta$, and free charge density $q$, of the form

\begin{equation}
\begin{array}{cc}
\sources\left(\locationN,t\right)=\sourcescon\mathrm{e}^{i\left(\kappa\locationN-\angvel t\right)}, & \sources=\strainso,\force,\charge,\end{array}
\end{equation}
where $\{\sourcescon\}$ are constant. These sources generate a longitudinal motion, $u\left(x,t\right)$, governed by the equations

\begin{equation}
\derivative{\stress}+\force=\derivativet{\momentum},\quad \derivative{\elecdisp}=\charge;
\label{eq:governingequation}
\end{equation}
here, $\stress$ is the Cauchy
stress, $\momentum$ is the linear momentum, and  $\elecdisp$ is the
electric displacement. The remaining Faraday equation for electric field $\elecfld$ is satisfied by setting $\elecfld = - \derivative{\elecpot}$, where $\elecpot$ is termed the electric potential. The microscopic kinetic and kinematic fields in the laminate are related via the constitutive relations 
\begin{equation}
\left(\begin{array}{c}
\stress\\
\elecdisp\\
\momentum
\end{array}\right)=\left(\begin{array}{ccc}
\stiffness & \Bb & 0\\
\Bb & -\Aa & 0\\
0 & 0 & \density
\end{array}\right)\left(\begin{array}{c}
\derivative{\disp}-\strainso\\
\derivative{\elecpot}\\
\derivativet{\disp}
\end{array}\right),
\label{eq:constPiezo}
\end{equation}
where $\stiffness$, $\Bb$, $A$ and $\rho$ are the stiffness, piezoelectric coefficient,  dielectric coefficient and the mass density, respectively. These properties are periodic, such that
\begin{equation}
\begin{array}{ccc}
\properties\left(\locationN+n\unit\right)=\properties\left(\locationN\right), & \properties=\stiffness,\Bb,\Aa,\density,\end{array}
\end{equation}
where $n\in\mathbb{Z}$; $l_{T} = \sum_{m=1}^{3} l_{m}$ and $l_{m}$ is the length of layer $m$, as shown in Figure \ref{fig:ThreeLayersExample}. This periodicity implies that the kinetic and the kinematic
fields are of the Bloch form 
\begin{equation}
\begin{aligned} & \begin{array}{ccc}
\hfield\left(\locationN,t\right)=\periodic{\hfield}\left(\locationN\right)e^{i\left(\kappa\locationN-\angvel t\right)}, & \periodic{\hfield}\left(\locationN+n\unit\right)=\periodic{\hfield}\left(\locationN\right), &\hfield=\stress,\elecdisp,\momentum,\disp,\elecpot.\end{array}
\end{aligned}
\end{equation}

The objective is to determine the effective constitutive relations, relating the macroscopic fields, which have the form  

\begin{equation}
\begin{array}{cc}
    \effective{\hfield}\left(\locationN,t\right)=\bar{\hfield} e^{i\left(\kappa\locationN-\angvel t\right)}, & \bar{\hfield} = \unit^{-1}\intop_{\unit}\periodic{\hfield}\left(\locationN\right)\mathrm{d\locationN}.
    \end{array}
\end{equation}


To this end, we first  consider the Bloch form of Eq.~\eqref{eq:constPiezo},
namely,
\begin{equation}
\left(\begin{array}{c}
\periodic{\stress}\left(\locs\right)\\
\periodic{\elecdisp}\left(\locs\right)\\
\periodic{\momentum}\left(\locs\right)
\end{array}\right)e^{i\left(\wavenum\locs-\angvel t\right)}=\left(\begin{array}{ccc}
\stiffness\left(\locs\right) & \trans{\Bb}\left(\locs\right) & 0\\
\Bb\left(\locs\right) & -\Aa\left(\locs\right) & 0\\
0 & 0 & \density\left(\locs\right)
\end{array}\right)\left(\begin{array}{c}
\derivative{\periodic{\disp}\left(\locs\right)}+i\wavenum\periodic{\disp}\left(\locs\right)-\periodic{\strainso}\left(\locs\right)\\
\derivative{\periodic{\elecpot}\left(\locs\right)}+i\wavenum\periodic{\elecpot}\left(\locs\right)\\
-i\angvel\periodic{\disp}\left(\locs\right)
\end{array}\right)e^{i\left(\wavenum\locs-\angvel t\right)}.
\label{eq:ConstBloch}
\end{equation}
To relate the mean parts of the kinetic and kinematic fields, we expand the periodic functions in Eq.~\eqref{eq:ConstBloch} in Fourier series, such that
\begin{equation}
\begin{array}{ccc}
    \periodic{\psi}\left(\locationN\right)=\sumG\coeffg{\psi}e^{i\reciprocal\locationN}, & \coeffg{\psi}=\unit^{-1}\intop_{\unit}\periodic{\psi}\left(\locationN\right)e^{-i\reciprocal\locationN}\mathrm{d\locationN},& \periodic{\psi} = \periodic{\hfield}, \properties,
    \label{eq:FourierSeries}
\end{array}
\end{equation}
where $\{\coeffg{\psi}\}$ are the Fourier coefficients of $\periodic{\psi}\left(\locationN\right)$; $\reciprocal=\frac{2\pi}{\unit}m$ and
$m\in\mathbb{Z}$.
Substituting Eq.~\eqref{eq:FourierSeries} into Eq.~\eqref{eq:ConstBloch} yields
\begin{equation}
\left[\sumG\left(\begin{array}{c}
\coeffg{\stress}\\
\coeffg{\elecdisp}\\
\coeffg{\momentum}
\end{array}\right)e^{i\reciprocal x}-\sumGtG\left(\begin{array}{ccc}
\coeffg{\stiffness} & \trans{\coeffg{\Bb}} & 0\\
\coeffg{\Bb} & -\coeffg{\Aa} & 0\\
0 & 0 & \coeffg{\density}
\end{array}\right)\left(\begin{array}{c}
i\kGt\coeffgt{\disp}-\coeffgt{\eta}\\
i\kGt\coeffgt{\elecpot}\\
-i\angvel\coeffgt{\disp}
\end{array}\right)e^{i\left(\reciprocal+\reciprocalt\right)x}\right]e^{i\left(\wavenum\locs-\angvel t\right)}=\left(\begin{array}{c}
0\\
0\\
0
\end{array}\right),
\label{eq:ConstFourier}
\end{equation}
where $\reciprocalt$ is defined in the same manner as $\reciprocal$ and Eq.~\eqref{eq:ConstFourier} contains every combination of $\reciprocalt$ and $\reciprocal$. Since Eq.~\eqref{eq:ConstFourier} holds for any value of $\locs$
and $t$, the sum in the square brackets must be zero. We multiply those sums by $e^{-i\reciprocal''x}$, and integrate the result over
the unit cell. Due to the orthogonality property of the Fourier series,
the only non-vanishing terms are those satisfying the condition $G''=\reciprocal+\reciprocalt$.
Hence, substituting $G'' = \reciprocal$ into the resultant equation yields
\begin{equation}
\left(\begin{array}{c}
\coeffg{\stress}\\
\coeffg{\elecdisp}\\
\coeffg{\momentum}
\end{array}\right)=\sumGt\left(\begin{array}{ccc}
\coeffggt{\stiffness} & \trans{\coeffggt{\Bb}} & 0\\
\coeffggt{\Bb} & -\coeffggt{\Aa} & 0\\
0 & 0 & \coeffggt{\density}
\end{array}\right)\left(\begin{array}{c}
i\kGt\coeffgt{\disp}-\coeffgt{\eta}\\
i\kGt\coeffgt{\elecpot}\\
-i\angvel\coeffgt{\disp}
\end{array}\right),\label{eq:Piezo-cosntFT}
\end{equation}
where $\ggt{\left(\circ\right)}G{G'}$ denotes the Fourier coefficient
of $\left(\circ\right)$ along the basis function $e^{i\left(\reciprocal-\reciprocalt\right)x}$
. In the same manner, Eq.~\eqref{eq:governingequation} reads 
\begin{equation}
\begin{aligned} & \left(\derivative{\periodic{\stress}}+i\wavenum\periodic{\stress}+\periodic{\force}\right)e^{i\left(\wavenum\locs-\angvel t\right)}=-i\angvel\periodic{\momentum}e^{i\left(\wavenum\locs-\angvel t\right)},\\
 & \left(\derivative{\periodic{\elecdisp}}+i\wavenum\periodic{\elecdisp}\right)e^{i\left(\wavenum\locs-\angvel t\right)}=\periodic{\charge}e^{i\left(\wavenum\locs-\angvel t\right)}.
\end{aligned}
\end{equation}
Once again, we expand the periodic parts to 
\begin{equation}
\begin{aligned} & \sumG\left[i\kG\coeffg{\stress}+\coeffg{\force}\right]e^{i\reciprocal x}e^{i\left(\wavenum\locs-\angvel t\right)}=-i\angvel\sumG\coeffg{\momentum}e^{i\reciprocal x}e^{i\left(\wavenum\locs-\angvel t\right)},\\
 & \sumG\left[i\kG\coeffg{\elecdisp}\right]e^{i\reciprocal x}e^{i\left(\wavenum\locs-\angvel t\right)}=\sumG\coeffg{\charge}e^{i\reciprocal x}e^{i\left(\wavenum\locs-\angvel t\right)}.
\end{aligned}
\end{equation}
A truncation to a finite number of plane waves $\numwaves$ is carried
out for implementational purposes, say by $-s\le m\le s$\footnote{The number of waves that was found sufficient for convergence in this work is $41$, see Appendix \ref{appendix2}.}, such that
$\numwaves=2s+1$. We multiply those sums by $e^{-i\reciprocal''x}$,
and integrate the result over the unit cell. Due to the orthogonality
property of the Fourier series, the only non-vanishing terms are those
satisfying the condition $G''=\reciprocal$. We can write for each
$\reciprocal$ 
\begin{equation}
\g{\opgov}^{\mathsf{T}}\g{\kineticv}=-\g{\forces},\label{eq:govG}
\end{equation}
where
\begin{equation}
\g{\opgov}=\left(\begin{array}{cc}
i\kG & 0\\
0 & i\kG\\
i\angvel & 0
\end{array}\right),\g{\kineticv}=\left(\begin{array}{c}
\coeffg{\stress}\\
\coeffg{\elecdisp}\\
\coeffg{\momentum}
\end{array}\right),\g{\forces}=\left(\begin{array}{c}
\coeffg{\force}\\
-\coeffg{\charge}
\end{array}\right).
\end{equation}
We assemble all the equations of each $\reciprocal$ into a single matrix system, namely,
\begin{equation}
\trans{\assembly{\opgov}}\assembly{\kineticv}=-\assembly{\forces},\label{eq:GovCona}
\end{equation}
where $\assembly{\forces}$ is a column vector of $2\numwaves$ components that assembles all the Fourier coefficients of $\coeffg{\force}$
and $\coeffg{\charge}$; $\assembly{\kineticv}$ is a column vector
of $3\numwaves$ components, which assembles all the Fourier coefficients
of $\coeffg{\stress},\coeffg{\elecdisp}$ and $\coeffg{\momentum}$;
and $\assembly{\opgov}$ is a $3\numwaves\times2\numwaves$ matrix that is composed from $3$ diagonal and $3$ zero matrices. For further details the reader is referred to 
Appendix \ref{appendix1}. In the same manner, $\assembly{\kineticv}$ is expressed by assembling the matrix equations of the constitutive relation in Eq.~\eqref{eq:Piezo-cosntFT}, namely,
\begin{equation}
\assembly{\kineticv}=\assembly{\prop}\left(\assembly{\opcon}\assembly{\kinematicv}-\assembly{\strainsov}\right),\label{eq:GovConb}
\end{equation}
where $\assembly{\prop}$ is a $3\numwaves\times3\numwaves$ matrix
that contains the Fourier components of the material properties.
Similarly to $\assembly{\opgov}$, $\assembly{\opcon}$ is a $3\numwaves\times2\numwaves$ matrix that is composed from 3 diagonal and 3 zero matrices; $\assembly{\kinematicv}$
is a column vector of $2\numwaves$ coefficients that assembles the
Fourier components of the electric potential the displacement;
and $\assembly{\strainsov}$ is a column vector of $3\numwaves$ components
that contains the Fourier coefficients of the eigenstrain. 
Extracting the average fields --- that are associated with $\reciprocal=0$ --- from Eq.~\eqref{eq:GovConb} yields
\begin{equation}
\average{\kineticv}\coloneqq\left(\begin{array}{c}
\average{\stress}\\
\average{\elecdisp}\\
\average{\momentum}
\end{array}\right)\equiv\left(\begin{array}{c}
\coeffz{\stress}\\
\coeffz{\elecdisp}\\
\coeffz{\momentum}
\end{array}\right)=\zero{\prop}\left\{ \zero{\opcon}\average{\kinematicv}-\average{\strainsov}\right\} +\reduced{\prop}\reduced{\opcon}\reduced{\kinematicv},\label{eq:constEFF}
\end{equation}
where $\zero{\opcon}$ and $\zero{\prop}$ are the parts of the matrices
that multiply the average fields and $\reduced{\left(\circ\right)}$ denotes the reduced matrix or vector without the $\reciprocal=0$ terms\footnote{Here $\zero{\prop}$ and $\reduced{\prop}$ is a $3\times3$ matrix and a $3\times\left(3\numwaves-3\right)$ matrix, respectively; $\zero{\opcon}$ and $\reduced{\opcon}$ is a $3\times2$ matrix and a $\left(3\numwaves-3\right)\times\left(2\numwaves-2\right)$
matrix, respectively; $\average{\kinematicv}$ and $\reduced{\kinematicv}$  is a vector with 2 components and a vector with $2\numwaves$, respectively, and $\average{\strainsov}$ is a vector with 3 components.}.

Our next step is to express the terms $\reduced{\kinematicv}$ using
$\average{\kinematicv}$ and $\average{\strainsov}$. To this end,
we first write Eq.~\eqref{eq:GovCona} in the form
\begin{equation}
\assembly{\trans{\opgov}}\assembly{\prop}\left(\assembly{\opcon}\assembly{\kinematicv}-\assembly{\strainsov}\right)=-\assembly{\forces}.\label{eq:ExpandGOV}
\end{equation}
Next, we rewrite Eq.~\eqref{eq:ExpandGOV} by separating the equations that do not include $\reciprocal=0$ and obtain
\begin{equation}
\reduced{\Q}\reduced{\kinematicv}=-\DL\left\{ \zero{\opcon}\average{\kinematicv}-\average{\strainsov}\right\} ,\label{eq:PiezoGovGne0}
\end{equation}
where $\reduced{\Q}$ is a square matrix whose entries multiply the average fields, and $\DL$ contains the terms of the
matrix $\trans{\opgov}\prop$ which multiply the average fields.
Next, the fluctuating terms of $\kinematicv$ can be expressed using its average value and the material properties, such that
\begin{equation}
\reduced{\kinematicv}=-\reduced{\Q}^{-1}\DL\left\{ \zero{\opcon}\average{\kinematicv}-\average{\strainsov}\right\} .\label{eq:PsiNeZero}
\end{equation}
Substituting Eq.~\eqref{eq:PsiNeZero} into Eq.~\eqref{eq:constEFF} yields

\begin{equation}
\average{\kineticv}=\zero{\prop}\left\{ \zero{\opcon}\average{\kinematicv}-\average{\strainsov}\right\} -\reduced{\prop}\reduced{\opcon}\reduced{\Q}^{-1}\DL\left\{ \zero{\opcon}\average{\kinematicv}-\average{\strainsov}\right\} .\label{eq:Compare}
\end{equation}
We can now  identify the effective properties that act on the macroscopic kinematic fields, i.e., on $\left\{ \zero{\opcon}\average{\kinematicv}-\average{\strainsov}\right\}$, namely, 
\begin{equation}
\left(\begin{array}{ccc}
\effectiveprop{\stiffness} & \dag{\effectiveprop{\Bb}} & \effectiveprop{\couplingS}\\
\effectiveprop{\Bb} & -\effectiveprop{\Aa} & \effectiveprop{\couplingW}\\
\dag{\effectiveprop{\couplingS}} & \dag{\effectiveprop{\couplingW}} & \effectiveprop{\density}
\end{array}\right):=\zero{\prop}-\reduced{\prop}\reduced{\opcon}\reduced{\Q}^{-1}\DL.\label{eq:W}
\end{equation} 
The effective properties \eqref{eq:W} are of the form reported in Refs.\ \cite{PernasSalomon2019JMPS,pernas2020fundamental,muhafra2021}: they are nonlocal in space and time, i.e., functions of $\kappa$ and $\angvel$; and include two pairs of adjoint properties---the Willis couplings $\tilde{S},\tilde{S}^\dagger$ and the electromomentum couplings  $\tilde{W},\tilde{W}^\dagger$---which  are absent in the microscopic relations. In this work, we focus on maximizing the electromomentum coupling at the long wavelength limit, namely at $\wavenum = 0$, in which case $\effectiveprop{\couplingS} = \dag{\effectiveprop{\couplingS}}$ and $\effectiveprop{\couplingW} = \dag{\effectiveprop{\couplingW}}$. We further restrict attention to the low frequency range, where the electromomentum coefficient is expected to depend linearly on the frequency \cite{PernasSalomon2019JMPS,rps20201wm}; Accordingly, we optimize $\tilde{W}$ at $2$ kHz. (Indeed, our calculations show that  in the range $0$ - $10$ kHz, $\tilde{W}$ grows linearly with the frequency.) Henceforth, the matrix in Eq.~\eqref{eq:W} will be denoted by $\Wmat$. In what follows, we analyze $\Wmat$, with the understanding that the component that is relevant to this work is $K_{23}$.

\section{Discrete material choice optimization}


\label{sec:DMO}
We begin by presenting a discrete material choice approach, where the base materials that compose the periodic structure are chosen from a predefined set of available materials given in Table \ref{tab:Table1}.
We focus on a laminate made of three constituents, optimize it over the choice of the materials, and evaluate simultaneously the length of each layer that leads to the optimal design, while the total length of the periodic cell is fixed to $3$\,mm.
Thanks to the use of standard base materials, the results of this design parameterization are directly applicable and can contribute significantly to practical implementation. 
As the optimization is based on discrete variables and the number of design variables is small, a genetic algorithm (GA) is used for solving the optimization problem.


The design variables are assigned to the choice of material in each of the three material layers, from a library of candidate materials, and to the length of each material phase.
This corresponds to the maximization problem
\begin{equation}
\begin{aligned}
& \underset{y_{j},l_1,l_2}{\text{maximize}}
& &|\effectiveprop{\couplingW}|, \\
& \text{subject to}
& & y_j \in [1,2,...10],\; j = 1,2,3,\\
& & & l_1 + l_2 \leq l_T, \\
& & & l_1, l_2 \geq 0, \\
\label{eq:DMO}
\end{aligned}
\end{equation}
where $y_j$ denotes the pointer to a discrete material that is a member of Table \ref{tab:Table1}.
For example, if $y_2 = 3$ then the second layer is Al$_{2}$O$_{3}$.
The length of the third layer is computed explicitly as the complementary of the total length, hence, it is not considered as a design variable. 
Note that for each minimal (negative) value of $\effectiveprop{\couplingW}$, there is an equivalent maximal (positive) value of $\effectiveprop{\couplingW}$ that can be designed by flipping the order of the structure, since the sign of $\effectiveprop{\couplingW}$ depends on the coordinate system \cite{rps20201wm}. Other operations can also yield equivalent designs with  the same absolute value of $\effectiveprop{\couplingW}$, as will be discussed in the next section.
Hence, our optimization implementation was to minimize the objective without referring to the absolute value, namely, to obtain the most negative electromomentum coupling.

In general, when using optimization methods, the optimization process might end up with a local minimum instead of a global one, and the outcome can strongly depend  on the initial starting point of the whole process. 
Hence, some form of verification is needed.
We define a confidence parameter to determine the threshold for a satisfactory solution.
The parameter is computed as follows

\begin{equation}
\label{eq:Conf}
\begin{aligned}
& E=\left(E_{1}E_{2}\right)^{0.5}, \\
& E_{1}=\frac{R_\text{{best}}}{R_\text{{total}}},\\
& E_{2}=e^{-2\frac{E_\text{{avg}}-E_\text{{best}}}{E_\text{{best}}}},
\end{aligned}
\end{equation}
where multiple runs have been considered;
$R_\text{{best}}$ is the number of runs that resulted in the optimal solution;
$R_\text{{total}}$ is the total number of runs;
$E_\text{{avg}}$ and $E_\text{{best}}$ are the average and optimal electromomentum coupling out of all the outcomes, respectively.
The minimum confidence required for ending the search for the optimal design depends on the number of runs, as presented
in Table \ref{tab:confidence}, together with the calculated confidence of the process.

\begin{table}[t!]
\begin{center}
\begin{tabular}{c| c c c c}
\hline
Material & $\stiffness$ [GPa] & $\density$ [kg/m$^3$] & $\Bb$ [C/m$^2$] & $\Aa$ [nF/m]\\
\hline
PZT4  & 115 & 7500 & 15.1 & 5.6\\
BaTiO$_3$  & 165 & 6020 & 3.64 & 0.97\\
Al$_2$O$_3$  & 300 & 3270 & 0 & 0.079\\
PMMA  & 3.3 & 1188 & 0 & 0.023\\
PZT , Navy type II PZT-5H  & 50 & 7500 & 29 & 30.1\\
LiNbO$_3$  & 200 & 4650 & 1.3 & 0.25\\
PMN-PT Single Crystal X2B  & 100 & 8100 & 200 & 53.1\\
Quartz  & 86 & 2650 & 0.17 & 0.04\\
Soft PZT HK1HD  & 50 & 8000 & 39 & 60.2\\
Hard PZT Type I  & 50 & 7900 & 15 & 11.5\\
\hline\hline
\end{tabular}
\caption{The library of candidate materials for the discrete optimization formulation \eqref{eq:DMO}.}
\label{tab:Table1} 
\end{center}
\end{table}

\begin{table}[t!]
\begin{center}
\begin{tabular}{c| c c}
\hline
Number of runs &Minimum confidence required & The calculated confidence \\
\hline
10 & 0.8 & 0.8323\\
20 & 0.5 & 0.8326\\
\hline\hline
\end{tabular}
\caption{Required and calculated confidence for ending the Genetic Algorithm search.}
\label{tab:confidence} 
\end{center}
\end{table}

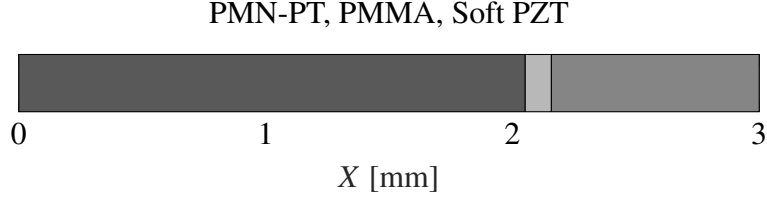
\begin{figure}[t!] 
\centering
%
%
\definecolor{mycolor1}{rgb}{0.35,0.35,0.35}%
\definecolor{mycolor2}{rgb}{0.72,0.72,0.72}%
\definecolor{mycolor3}{rgb}{0.52,0.52,0.52}%
\begin{tikzpicture}

\begin{axis}[%
width=3.875in,
height=0.3in,
at={(0.65in,0.673in)},
scale only axis,
xmin=0,
xmax=3,
xtick={0, 1, 2, 3},
xlabel style={font=\color{white!15!black}},
xlabel={$X$ [mm]},
ymin=0,
ymax=1,
ytick={\empty},
axis background/.style={fill=white},
title={PMN-PT, PMMA, Soft PZT}
]
\draw[fill=mycolor1, draw=black] (axis cs:0,0) rectangle (axis cs:2.052,1);
\draw[fill=mycolor2, draw=black] (axis cs:2.052,0) rectangle (axis cs:2.158,1);
\draw[fill=mycolor3, draw=black] (axis cs:2.158,0) rectangle (axis cs:3,1);
\end{axis}
\end{tikzpicture}%
\caption{The optimal design according to the discrete material choice approach, found by the GA. Each color represents a different material. The lengths of the layers are $2.052$ mm, $0.106$ mm and $0.842$ mm and $|\effectiveprop{\couplingW}| = 6.312\cdot10^{-7}$ Cs/m$^{3}$.}
\label{fig:DMO}
\end{figure}

Our results show that the optimal structure is $2.052$ mm PMN-PT Single Crystal X2B--$0.106$ mm PMMA--$0.842$ mm Soft PZT HK1HD, (Figure \ref{fig:DMO}), yielding  $\tilde{W}=6.312\cdot10^{-7}$ Cs/m$^{3}$, which is one order of magnitude larger than the  electromomentum coefficient of the reference laminate. 
We observe that the optimizer provided a laminate with high contrast between the properties of its constituents: PMMA has the smallest value of each one of the electromechanical properties; PMN-PT Single Crystal
X2B has the largest piezoelectric coefficient and mass density; PZT HK1HD has the largest dielectric coefficient and second-largest mass density. 
The asymmetry is  not only in the material properties but also in the geometry: the middle layer is very thin with respect to the other two layers. The above features reoccur in subsequent sections.

\section{Gradient-based optimization}
\label{sec:GradientBased}

The efficiency of genetic algorithms strongly depends  on the number of design
variables,  exhibiting a sharp decline when the number of design
variables increases.
Even though the one-dimensional problem that we address does not require a large number of design variables, the capability to optimize with a large number of variables is desirable in general; for example, in  extensions of the current work to higher-dimensional problems.  
In such cases, gradient-based optimization is more suitable, and 
this section is dedicated to  optimizations using continuous variables that is solved by such gradient-based algorithms.
In its most general form, the optimization problem includes design variables that govern the material properties and the geometry of each of the three material phases.
An essential building block is the derivation of the analytical sensitivity analysis of the objective function with respect to the design variables, which we carry out in the sequel. 


The general problem consists of finding the optimal material properties and the optimal length of each layer. 
This leads to the following maximization problem
\begin{equation}
\label{eq:formulationFMO}
\begin{aligned}
& \underset{z_{mn},l_1,l_2}{\text{maximize}}
& &|\effectiveprop{\couplingW}|,\\
& \text{subject to}
& & l_1 + l_2 \leq l_T, \\
& & & l_1, l_2 \geq 0, \\
& & & z_{mL} \leq z_{mn} \leq  z_{mU} \:\:\:\: \forall m=1 \ldots 4,n=1 \ldots 3,
\end{aligned}
\end{equation}
where $z_{mn}$ denotes the material property $m$, i.e.,  $\Bb$, $\Aa$, $\stiffness$, $\rho$, in this order, of layer $n$.
For example, $z_{23}$ is the dielectric coefficient of the third layer.
The notation $z_{mL}$ and $z_{mU}$ represents the lower and upper limits of  property $m$. 
Each material property is optimized independently, thus relaxing the constraint on the relation between the various material properties.
The purpose of this FMO \cite{zowe1997free} is twofold:
first, it delivers a theoretical upper bound on the electromomentum
coupling;
second, it provides guidelines on how to compose a laminate that maximizes $\effectiveprop{\couplingW}$. 

In order to obtain results with some correlation to realistic material properties, the design variables are limited in-between the extreme properties among the real materials considered in Section \ref{sec:DMO}, as presented in Table \ref{tab:Table4}.  
The material properties were normalized with respect to the same limits. The normalized properties are computed as follows 
\begin{equation}
    \chi_{m}=\frac{\chi_{phys}-\chi_{L}}{\chi_{U}-\chi_{L}},
\end{equation}
where $\chi_{m}$ and $\chi_{phys}$ are the normalized and the physical
property, respectively; and $\chi_{L}$ and $\chi_{U}$ are the lower and
upper limits, respectively. 

\begin{table}[t!]
\begin{center}
\begin{tabular}{c| c c }
\hline
  & Upper limits & Lower limits\\
\hline
B [C/m$^2$] & 200 & 0\\
A [nF/m] & 60.2 & 0.023\\
C [GPa] & 300 & 3.3\\
$\rho$ [kg/m$^3$] & 1188 & 8100\\
\hline\hline
\end{tabular}
\caption{Upper and lower limits for the material properties in Eq.~\eqref{eq:formulationFMO}.}
\label{tab:Table4}
\end{center}
\end{table}

\subsection{Sensitivity analysis}\label{sec:SA_contin}
We carry out a differentiation of the objective function with respect to each one of the design variables, i.e.,  the lengths of the first two layers and the material properties of each layer.
Considering first  the length variables, the derivative of Eq.~\eqref{eq:W} using the product rule yields
\begin{equation}
\frac{\partial\Wmat}{\partial l_{n}}=-\left[\frac{\partial\reduced{\prop}}{\partial
l_{n}}\reduced{\opcon}\reduced{\Q}^{-1}\DL+\reduced{\prop}\reduced{\opcon}\frac{{\partial\reduced{\Q}}^{-1}}{\partial
l_{n}}\DL+\reduced{\prop}\reduced{\opcon}\reduced{\Q}^{-1}\frac{\partial\DL}{\partial
l_{n}}\right].
\label{eq:bigder}
\end{equation}
Since the PWE expands the governing- and constitutive equations in Fourier series, the entries in each matrix listed in Eq.~\eqref{eq:bigder} are basically Fourier coefficients, as given in Eq.~\eqref{eq:FourierSeries}. Thus, differentiating $\coeffg{\properties}$ and following the same procedure yields the derivatives that we are looking for, namely $\frac{\partial\reduced{\prop}}{\partial
l_{n}}, \frac{{\partial\reduced{\Q}}}{\partial
l_{n}}$ and $\frac{\partial\DL}{\partial
l_{n}}$. 
Since we are analyzing trilayer laminates, the integral is divided into three parts, namely, 
\begin{equation}
\coeffg{\properties}=\unit^{-1}\left[\intop_{0}^{l_{1}}\properties_{1}\mathrm{e}^{-iG\locationN}\mathrm{d\locationN}+\intop_{l_{1}}^{l_{2}}\properties_{2}\mathrm{e}^{-iG\locationN}\mathrm{d\locationN}+\intop_{l_{2}}^{l_{3}}\properties_{3}\mathrm{e}^{-iG\locationN}\mathrm{d\locationN}\right]. 
\label{eq: FourPhrase}
\end{equation}
Then, following the product rule, the derivative is expressed as
\begin{align}
\frac{\partial \coeffg{\properties}}{\partial l_{1}} & =\frac{1}{\unit}\left\{ \properties_{1}e^{-iG(l_{1})}+\properties_{2}\left[e^{-iG(l_{1}+l_{2})}-e^{-iG(l_{1})}\right]+\properties_{3}\left[e^{-iG(l_{1}+l_{2}+l_{3})}-e^{-iG(l_{1}+l_{2})}\right]\right\},\\
\frac{\partial \coeffg{\properties}}{\partial l_{2}} & =\frac{1}{\unit}\left\{ \properties_{2}e^{-iG(l_{1}+l_{2})}+\properties_{3}\left[e^{-iG(l_{1}+l_{2}+l_{3})}-e^{-iG(l_{1}+l_{2})}\right]\right\},\\
\frac{\partial \coeffg{\properties}}{\partial l_{3}} & = \frac{1}{\unit}\left\{ \properties_{3}e^{-iG(l_{1}+l_{2}+l_{3})}]\right\} . \label{eq:div}
\end{align}
The chain rule yields
\begin{equation}
\frac{d(\cdot)}{dl_{j}}  = \frac{\partial (\cdot)}{\partial l_{j}} + \frac{\partial (\cdot)}{\partial l_{3}}\frac{\partial l_{3}}{\partial l_{j}} \:\:\:\:, \quad j=1,2,\\
\end{equation}
and we recall that  total length of the periodic cell is fixed,  hence
\begin{equation}
\frac{d(\cdot)}{dl_{j}}  = \frac{\partial (\cdot)}{\partial l_{j}} - \frac{\partial (\cdot)}{\partial l_{3}},
\end{equation}
where $(\cdot)$ denotes the matrices $\reduced{\prop}$,  $\reduced{\Q}$ and $\DL$. Now that we have the derivatives above, we can follow the same process as described in Section \ref{chap:PWE}, and obtain the desired derivatives in the same manner as in computing $\reduced{\prop}$,  $\reduced{\Q}$ and $\DL$.
Note that according to Eq.~\eqref{eq:bigder}, one must compute
the derivative of $\Q_{s}^{-1}$, and this is carried out using the
relation
\begin{equation}
\frac{\partial\Q_{s}^{-1}}{\partial l_{j}}=-\Q_{s}^{-1}\cdot\frac{\partial\Q_{s}}{\partial l_{j}}\cdot\Q_{s}^{-1}.
\end{equation}

The sensitivities with respect to material properties are derived similarly to the procedure outlined above for the length variables. 
Analogously to Eq.~\eqref{eq:bigder}, the sensitivity with respect to any material property is expressed as
\begin{equation}
\frac{\partial\Wmat}{\partial z_{mn}}=-\left[\frac{\partial\reduced{\prop}}{\partial z_{mn}}\reduced{\opcon}\reduced{\Q}^{-1}\DL+\reduced{\prop}\reduced{\opcon}\frac{\partial\reduced{\Q}^{-1}}{\partial z_{mn}}\DL+\reduced{\prop}\reduced{\opcon}\reduced{\Q}^{-1}\frac{\partial\DL}{\partial z_{mn}}\right].
\label{eq:bigder2}
\end{equation}
Again, the entries of the matrices are Fourier coefficients, therefore the derivative of Eq.~\eqref{eq: FourPhrase} with respect to the material properties is 
\begin{align}\frac{\partial \coeffg{\properties}}{\partial z_{m1}} & =\frac{-1}{iG\unit}\left[e^{-iG(l_{1})}-1\right],\\
\frac{\partial \coeffg{\properties}}{\partial z_{m2}} & =\frac{-1}{iG\unit}\left[e^{-iG(l_{1}+l_{2})}-e^{-iGl_{1}}\right],\\
\frac{\partial \coeffg{\properties}}{\partial z_{m3}} & =\frac{-1}{iG\unit}\left[e^{-iG(l_{1}+l_{2}+l_{3})}-e^{-iG(l_{1}+l_{2})}\right].
\label{defM}
\end{align}
Henceforth, a procedure analogous to the PWE is carried out to obtain the required sensitivities, namely $\frac{\partial\reduced{\prop}}{\partial z_{mn}}, \frac{\partial\reduced{\Q}}{\partial z_{mn}}$ and $\frac{\partial\DL}{\partial z_{mn}}$. 
Note that we have compared the evaluation of the analytical expressions for the derivatives with numerical derivatives that were obtained by finite differences. This comparison  yielded an excellent agreement, thereby verifying the correctness of our sensitivity analysis.

\subsection{Test case: optimization over the lengths}
\label{sec:result_optLen}

We begin our numerical investigation with a benchmark problem of length optimization;  this simple problem allows us to graphically verify our optimization method.

\begin{figure}[t!] 
\centering
\subfloat[]{\label{fig:cont1} 
\includegraphics[width=0.5\textwidth]{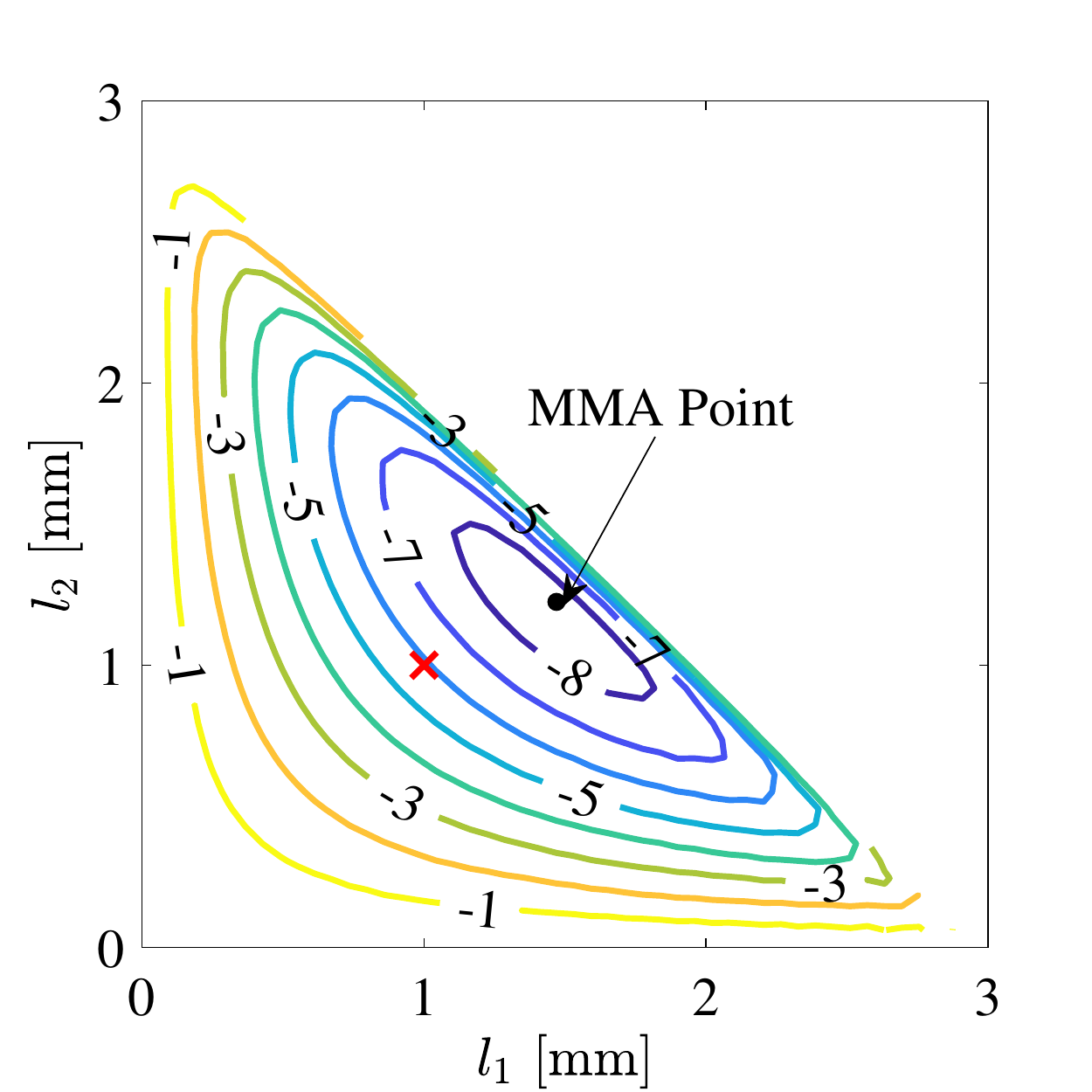}}
\subfloat[]{\label{fig:optlen1}
\includegraphics[width=0.5\textwidth]{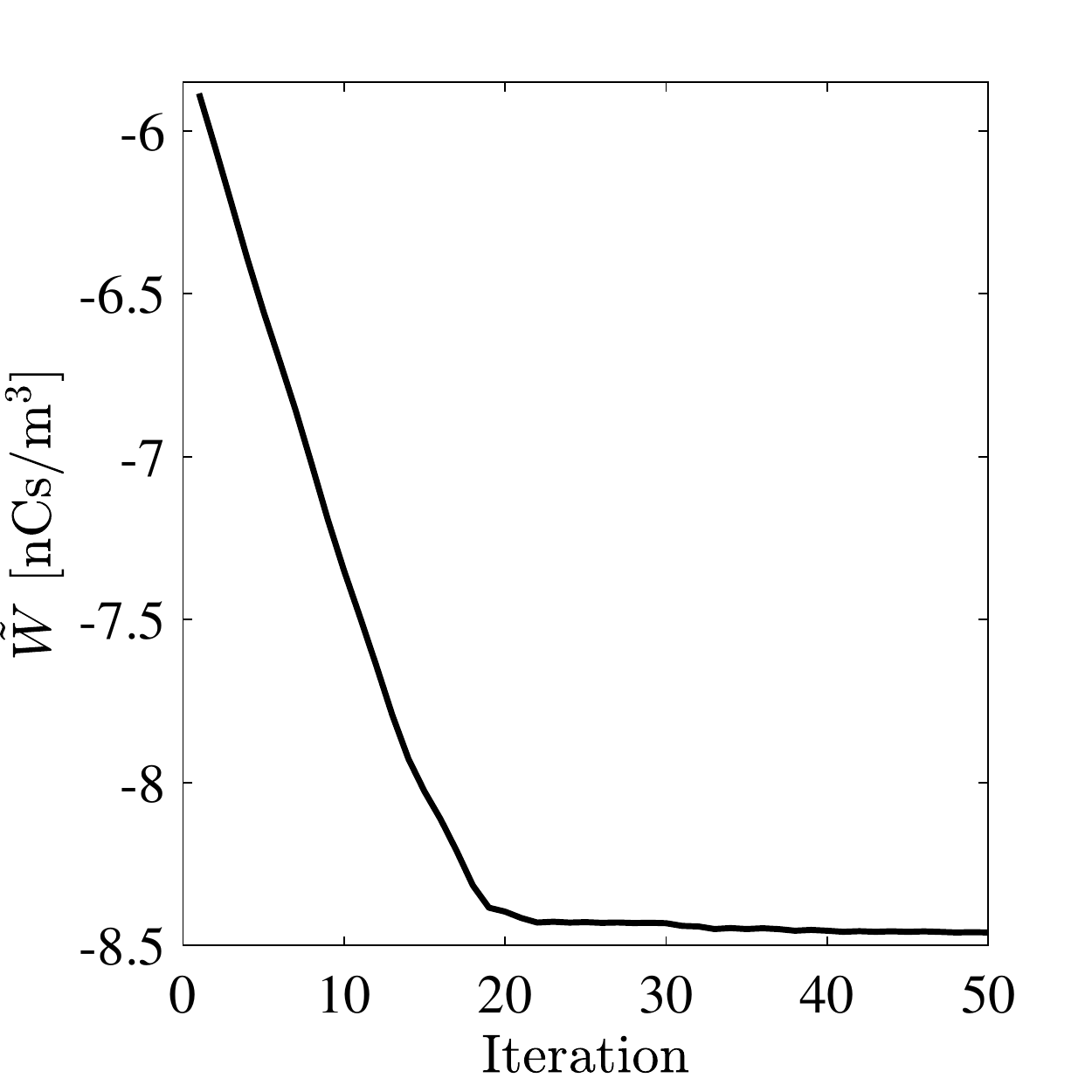}}\\
\subfloat[]{\label{fig:cont2}
\includegraphics[width=0.5\textwidth]{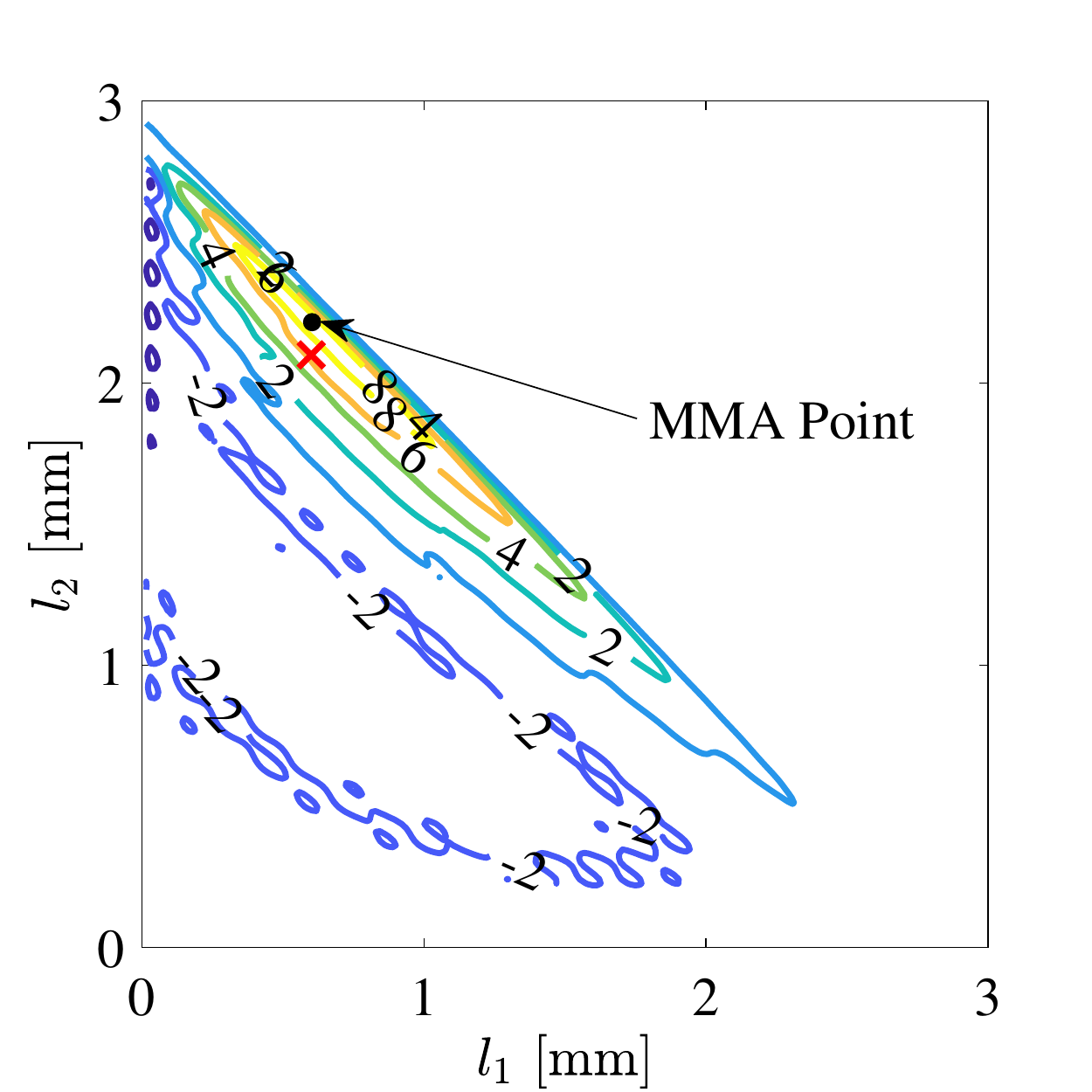}} 
\subfloat[]{\label{fig:optlen2}
\includegraphics[width=0.5\textwidth]{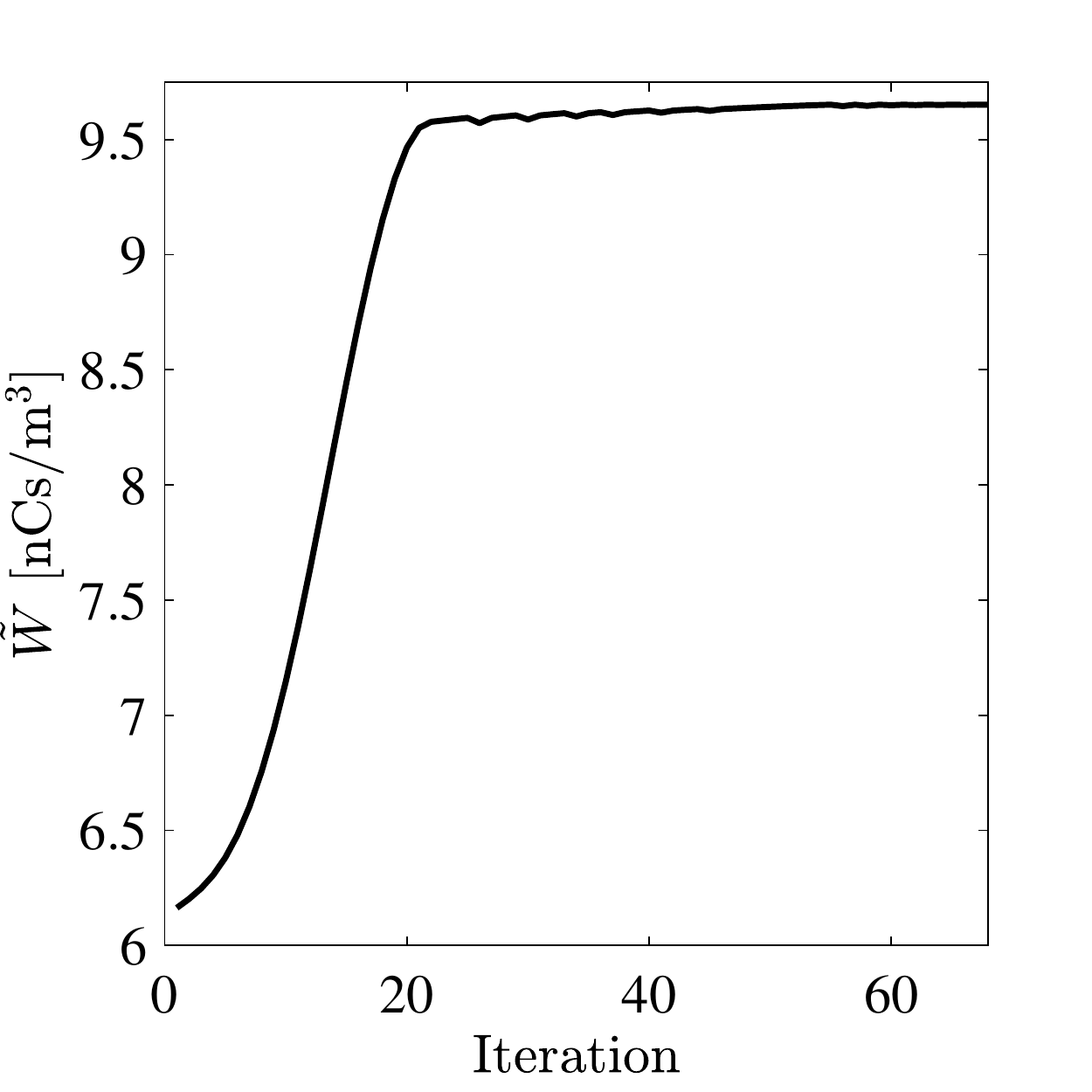}} 
\caption{Results of the length optimization. Panels (a) and (b) present the optimal solution for the first composition: (a) Graphical solution; (b) Convergence of the gradient-based optimization. Panels (c) and (d) present the optimal solution for the second composition: (c) Graphical solution; (d) Convergence of the gradient-based optimization. In both contours the initial (optimal) design is marked with red cross (black dot), and the objective function is multiplied by $10^9$.}
\label{fig:optlen}
\end{figure}

We consider two triplets of materials. The first triplet consists of the materials PZT4, BaTiO$_{3}$ and Al$_{2}$O$_{3}$. 
The graphical solution is presented in Figure \ref{fig:cont1}, where we
observe that the optimal design is located near the diagonal.
This means that the third layer should be thin in order
to maximize the absolute value of $\effectiveprop{\couplingW}$, which is
found to be $8.462\cdot10^{-9}$ Cs/m$^{3}$. 
Solving the problem by a numerical optimization scheme using the method of moving asymptotes  \cite{svanberg1987method}, we obtain the same result with the iterative convergence as plotted in Figure \ref{fig:optlen1}.
The optimal solution is characterized by the material distribution of $l_1=1.38$ mm, $l_2 = 1.28$ mm and $l_3 = 0.3979$ mm.

The second triplet consists of the materials Al$_{2}$O$_{3}$, BaTiO$_{3}$ and PMMA, which corresponds to the reference laminate when all layers have the same length.
Figure \ref{fig:cont2} shows the graphical solution, where we observe that the optimal design is located near the diagonal.
The gradient-based optimization  achieves  the same result, $\effectiveprop{\couplingW} = 9.52\cdot10^{-9}$ Cs/m$^{3}$, as indicated in Figure \ref{fig:optlen2}, which shows the iterative convergence of the gradient-based optimization. This result reflects a fourfold improvement with respect to the reference laminate, achieved solely by changing its geometry. Specifically, this optimal design is the lengths $l_1=0.69$ mm, $l_2 = 2.13$ mm and $l_3 = 0.19$ mm, for layers one to three, respectively. 
Both results are summarized in Table \ref{tab:length_opt}. 
We observe that the optimal designs share a common feature: a thin layer with the lowest mass density and piezoelectric coefficient. 
This is a repeating feature throughout the following optimization problems.

The graphical solution in Figure \ref{fig:cont2} shows that the optimization problem with this set of materials exhibits several local minima.
Using a gradient-based method, this might lead to results that depend heavily on the initial structure.
In order to overcome this problem, the optimization process has been divided into two steps. The first step seeks the optimal region by calculating the objective function at 20 random points inside the feasible region. 
Once the best starting point is found, in the second step we use the gradient-based optimization to steer the design towards a nearby minimum point.

As inferred from Eq.~\eqref{eq:formulationFMO}, the length of the third layer is not a design variable.
This may hamper convergence because the maximum change in the length of the third layer is related to the design variables,  which are the lengths of the two other layers.
For example, in cases where the optimal solution is located near the diagonal, the length of the third layer is small compared to the first and the second layers.
Then a design change that is suitable for the design variables, might be too liberal for the third layer and divergence may occur.
Therefore, conservative values of a move limit on the change of the design variables were imposed.
In all examples, a maximum move limit of 0.03 mm was enforced. \par

\newcommand{\centered}[1]{\begin{tabular}{l} #1 \end{tabular}}
\begin{center}
\begin{table}[t]
\begin{tabular}{@{}c@{}@{}c@{}@{}c@{}}
\hline
  \centered{ Composition} &
  \centered{
    $\effectiveprop{\couplingW}$: \text{Equally divided design } [Cs/m$^3$] }& 
  \centered{
    $\effectiveprop{\couplingW}$: \text{Optimal design } [Cs/m$^3$] } \\
  \hline
  \centered{
    PZT4-BaTiO$_3$-Al$_{2}$O$_{3}$ } &
  \centered{
%
%
\definecolor{mycolor1}{rgb}{0.35,0.35,0.35}%
\definecolor{mycolor2}{rgb}{0.72,0.72,0.72}%
\definecolor{mycolor3}{rgb}{0.52,0.52,0.52}%
\begin{tikzpicture}

\begin{axis}[%
width=2in,
height=0.3in,
at={(0.65in,0.675in)},
scale only axis,
xmin=0,
xmax=3,
xtick={0, 1, 2, 3},
xlabel style={font=\color{white!15!black}},
xlabel={$X$ [mm]},
ymin=0,
ymax=1,
ytick={\empty},
axis background/.style={fill=white},
title style={font=\bfseries},
title={$\tilde{W}=-5.88\cdot10^{-9}$}
]
\draw[fill=mycolor2, draw=black] (axis cs:0,0) rectangle (axis cs:1,1);
\draw[fill=mycolor3, draw=black] (axis cs:1,0) rectangle (axis cs:2,1);
\draw[fill=mycolor1, draw=black] (axis cs:2,0) rectangle (axis cs:3,1);
\end{axis}
\end{tikzpicture}%
 } &
  \centered{
%
%
\definecolor{mycolor1}{rgb}{0.35,0.35,0.35}%
\definecolor{mycolor2}{rgb}{0.72,0.72,0.72}%
\definecolor{mycolor3}{rgb}{0.52,0.52,0.52}%
\begin{tikzpicture}

\begin{axis}[%
width=2in,
height=0.3in,
at={(0.65in,0.675in)},
scale only axis,
xmin=0,
xmax=3,
xtick={0, 1, 2, 3},
xlabel style={font=\color{white!15!black}},
xlabel={$X$ [mm]},
ymin=0,
ymax=1,
ytick={\empty},
axis background/.style={fill=white},
title style={font=\bfseries},
title={$\tilde{W}=-8.46\cdot10^{-9}$}
]
\draw[fill=mycolor2, draw=black] (axis cs:0,0) rectangle (axis cs:1.384,1);
\draw[fill=mycolor3, draw=black] (axis cs:1.384,0) rectangle (axis cs:2.6021,1);
\draw[fill=mycolor1, draw=black] (axis cs:2.6021,0) rectangle (axis cs:3,1);
\end{axis}
\end{tikzpicture}%
  } \\
    \hline
    
      \centered{
    Al$_{2}$O$_{3}$-BaTiO$_3$-PMMA } &
  \centered{
%
%
\definecolor{mycolor1}{rgb}{0.35,0.35,0.35}%
\definecolor{mycolor2}{rgb}{0.82,0.82,0.82}%
\definecolor{mycolor3}{rgb}{0.52,0.52,0.52}%
%
\begin{tikzpicture}

\begin{axis}[%
width=2in,
height=0.3in,
at={(0.65in,0.675in)},
scale only axis,
xmin=0,
xmax=3,
xtick={0, 1, 2, 3},
xlabel style={font=\color{white!15!black}},
xlabel={$X$ [mm]},
ymin=0,
ymax=1,
ytick={\empty},
axis background/.style={fill=white},
title style={font=\bfseries},
title={$\tilde{W}=-2.34\cdot10^{-9}$}
]
\draw[fill=mycolor1, draw=black] (axis cs:0,0) rectangle (axis cs:1,1);
\draw[fill=mycolor3, draw=black] (axis cs:1,0) rectangle (axis cs:2,1);
\draw[fill=mycolor2, draw=black] (axis cs:2,0) rectangle (axis cs:3,1);
\end{axis}
\end{tikzpicture}%
 } &
  \centered{
%
%
\definecolor{mycolor1}{rgb}{0.35,0.35,0.35}%
\definecolor{mycolor2}{rgb}{0.82,0.82,0.82}%
\definecolor{mycolor3}{rgb}{0.52,0.52,0.52}%
%
\begin{tikzpicture}

\begin{axis}[%
width=2in,
height=0.3in,
at={(0.65in,0.675in)},
scale only axis,
xmin=0,
xmax=3,
xtick={0, 1, 2, 3},
xlabel style={font=\color{white!15!black}},
xlabel={$X$ [mm]},
ymin=0,
ymax=1,
ytick={\empty},
axis background/.style={fill=white},
title style={font=\bfseries},
title={$\tilde{W}=9.52\cdot10^{-9}$}
]
\draw[fill=mycolor1, draw=black] (axis cs:0,0) rectangle (axis cs:0.6855,1);
\draw[fill=mycolor3, draw=black] (axis cs:0.6855,0) rectangle (axis cs:2.8143,1);
\draw[fill=mycolor2, draw=black] (axis cs:2.8143,0) rectangle (axis cs:3,1);
\end{axis}
\end{tikzpicture}%
  } \\
    \hline
    \hline
\end{tabular}
\caption{Results of length optimization using gradient-based optimization. The optimal lengths of the layers in the first composition are 1.38 mm, 1.28 mm and 0.3979 mm, and for the second composition they are 0.69 mm, 2.13 mm and 0.19 mm. Each color represents a different material.
Both optimal designs exhibit a structure with one relatively thin layer.} 
\label{tab:length_opt} 
\end{table}
\end{center}

\subsection{Optimization over different material properties}
\label{sec:res_subsets}

Having verified our gradient-based optimization using a test case, 
we proceed to investigate optimization problems whose design variables are chosen material properties, where the rest of the properties are set to the properties of  Navy type II PZT-5H. 

\begin{figure}[t] 
\subfloat[]{\label{fig:Brho1}
\resizebox{0.5\textwidth}{!}{
%
%
\begin{tikzpicture}

\begin{axis}[%
width=3.353in,
height=0.6in,
at={(0.562in,0.675in)},
scale only axis,
point meta min=0,
point meta max=1,
colormap/jet,
separate axis lines,
every outer x axis line/.append style={black},
every x tick label/.append style={font=\color{black}, font=\fontsize{18}{18}\selectfont},
every x tick/.append style={black},
xmin=0,
xmax=3,
xtick={0, 1, 2, 3},
xlabel={\fontsize{18}{18}\selectfont $X$ [mm]},
every outer y axis line/.append style={black},
every y tick label/.append style={font=\color{black}, font=\fontsize{18}{18}\selectfont},
every y tick/.append style={black},
ymin=0,
ymax=1,
ytick={0.25,0.75},
yticklabels={{$\rho$},{$\text{\it{} B}$}},
axis background/.style={fill=white},
colorbar
]

\addplot[area legend, table/row sep=crcr, patch, patch type=rectangle, shader=flat corner, draw=black, forget plot, patch table with point meta={%
0	1	2	3	0\\
4	5	6	7	0\\
8	9	10	11	1\\
12	13	14	15	1\\
16	17	18	19	0\\
20	21	22	23	0.4813\\
}]
table[row sep=crcr] {%
x	y\\
0	0\\
1	0\\
1	0.5\\
0	0.5\\
0	0.5\\
1	0.5\\
1	1\\
0	1\\
1	0\\
2	0\\
2	0.5\\
1	0.5\\
1	0.5\\
2	0.5\\
2	1\\
1	1\\
2	0\\
3	0\\
3	0.5\\
2	0.5\\
2	0.5\\
3	0.5\\
3	1\\
2	1\\
};
\end{axis}
\end{tikzpicture}%
}}
\subfloat[]{\label{fig:Brho2}
\resizebox{0.5\textwidth}{!}{
%
%
\begin{tikzpicture}

\begin{axis}[%
width=3.353in,
height=0.6in,
at={(0.562in,0.675in)},
scale only axis,
point meta min=0,
point meta max=1,
colormap/jet,
separate axis lines,
every outer x axis line/.append style={black},
every x tick label/.append style={font=\color{black}, font=\fontsize{18}{18}\selectfont},
every x tick/.append style={black},
xmin=0,
xmax=3,
xtick={0, 1, 2, 3},
xlabel={\fontsize{18}{18}\selectfont $X$ [mm]},
every outer y axis line/.append style={black},
every y tick label/.append style={font=\color{black},font=\fontsize{18}{18}\selectfont},
every y tick/.append style={black},
ymin=0,
ymax=1,
ytick={0.25,0.75},
yticklabels={{$\rho$},{$\text{\it{} B}$}},
axis background/.style={fill=white},
colorbar
]

\addplot[area legend, table/row sep=crcr, patch, patch type=rectangle, shader=flat corner, draw=black, forget plot, patch table with point meta={%
0	1	2	3	1\\
4	5	6	7	0.4813\\
8	9	10	11	0\\
12	13	14	15	1\\
16	17	18	19	1\\
20	21	22	23	0\\
}]
table[row sep=crcr] {%
x	y\\
0	0\\
1	0\\
1	0.5\\
0	0.5\\
0	0.5\\
1	0.5\\
1	1\\
0	1\\
1	0\\
2	0\\
2	0.5\\
1	0.5\\
1	0.5\\
2	0.5\\
2	1\\
1	1\\
2	0\\
3	0\\
3	0.5\\
2	0.5\\
2	0.5\\
3	0.5\\
3	1\\
2	1\\
};
\end{axis}
\end{tikzpicture}%
}}\\
\subfloat[]{\label{fig:Brho3}
\resizebox{0.5\textwidth}{!}{
%
%
\begin{tikzpicture}

\begin{axis}[%
width=3.353in,
height=0.6in,
at={(0.562in,0.675in)},
scale only axis,
point meta min=0,
point meta max=1,
colormap/jet,
separate axis lines,
every outer x axis line/.append style={black},
every x tick label/.append style={font=\color{black}, font=\fontsize{18}{18}\selectfont},
every x tick/.append style={black},
xmin=0,
xmax=3,
xtick={0, 1, 2, 3},
xlabel={\fontsize{18}{18}\selectfont $X$ [mm]},
every outer y axis line/.append style={black},
every y tick label/.append style={font=\color{black}, font=\fontsize{18}{18}\selectfont},
every y tick/.append style={black},
ymin=0,
ymax=1,
ytick={0.25,0.75},
yticklabels={{$\rho$},{$\text{\it{} B}$}},
axis background/.style={fill=white},
colorbar
]

\addplot[area legend, table/row sep=crcr, patch, patch type=rectangle, shader=flat corner, draw=black, forget plot, patch table with point meta={%
0	1	2	3	0\\
4	5	6	7	0.4813\\
8	9	10	11	1\\
12	13	14	15	1\\
16	17	18	19	0\\
20	21	22	23	0\\
}]
table[row sep=crcr] {%
x	y\\
0	0\\
1	0\\
1	0.5\\
0	0.5\\
0	0.5\\
1	0.5\\
1	1\\
0	1\\
1	0\\
2	0\\
2	0.5\\
1	0.5\\
1	0.5\\
2	0.5\\
2	1\\
1	1\\
2	0\\
3	0\\
3	0.5\\
2	0.5\\
2	0.5\\
3	0.5\\
3	1\\
2	1\\
};
\end{axis}
\end{tikzpicture}%
}}
\subfloat[]{\label{fig:Brho4}
\resizebox{0.5\textwidth}{!}{
%
%
\begin{tikzpicture}

\begin{axis}[%
width=3.353in,
height=0.6in,
at={(0.562in,0.675in)},
scale only axis,
point meta min=0,
point meta max=1,
colormap/jet,
separate axis lines,
every outer x axis line/.append style={black},
every x tick label/.append style={font=\color{black}, font=\fontsize{18}{18}\selectfont},
every x tick/.append style={black},
xmin=0,
xmax=3,
xtick={0, 1, 2, 3},
xlabel={\fontsize{18}{18}\selectfont $X$ [mm]},
every outer y axis line/.append style={black},
every y tick label/.append style={font=\color{black}, font=\fontsize{18}{18}\selectfont},
every y tick/.append style={black},
ymin=0,
ymax=1,
ytick={0.25,0.75},
yticklabels={{$\rho$},{$\text{\it{} B}$}},
axis background/.style={fill=white},
colorbar
]

\addplot[area legend, table/row sep=crcr, patch, patch type=rectangle, shader=flat corner, draw=black, forget plot, patch table with point meta={%
0	1	2	3	1\\
4	5	6	7	0\\
8	9	10	11	0\\
12	13	14	15	1\\
16	17	18	19	1\\
20	21	22	23	0.4813\\
}]
table[row sep=crcr] {%
x	y\\
0	0\\
1	0\\
1	0.5\\
0	0.5\\
0	0.5\\
1	0.5\\
1	1\\
0	1\\
1	0\\
2	0\\
2	0.5\\
1	0.5\\
1	0.5\\
2	0.5\\
2	1\\
1	1\\
2	0\\
3	0\\
3	0.5\\
2	0.5\\
2	0.5\\
3	0.5\\
3	1\\
2	1\\
};
\end{axis}
\end{tikzpicture}%
}}
\caption{Four equivalent optimal designs as an outcome of the FMO over the piezoelectric and the mass density properties. Panels (a) and (b) share the same sign of the electromomentum coupling, while panels (c) and (d) have the opposite sign with the same magnitude. The properties were color-scaled as yellow (blue) color denotes the upper (lower) limit value. The optimal $|\effectiveprop{\couplingW}|$ for all four designs is $5.687\cdot10^{-6}$ Cs/m$^{3}$.}
\label{fig:BrhoB}
\end{figure}
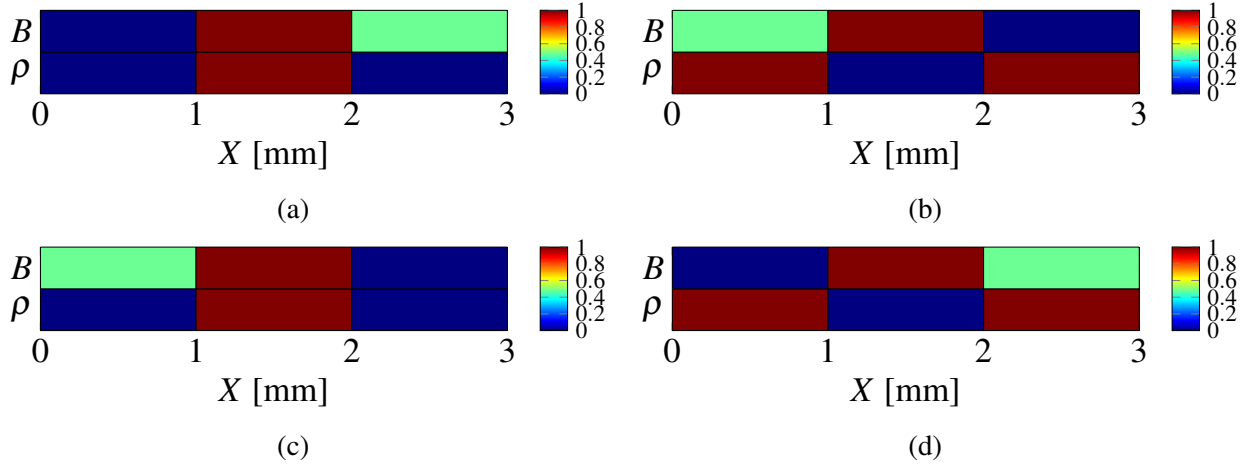

We begin by considering the design space of the piezoelectric coefficient and  mass density of each layer, yielding six design variables. 
For this case, the optimal electromomentum coefficient is $5.687\cdot10^{-6}$ Cs/m$^{3}$, which is three orders of magnitude greater than than that of the reference laminate. This value is achieved by four equivalent designs, as displayed in Figure \ref{fig:BrhoB}; the properties were color-scaled as yellow (blue) color denotes the upper (lower) limit value. The same color scaling will be used in the subsequent results. 
The optimal design exhibits a large contrast between the material properties, as 5 out of 6 properties attain the limit values, and one remaining piezoelectric coefficient attains the intermediate value $96.25$ C/m$^{2}$.
This result qualitatively agrees with the insights of \citet{rps20201wm}, who carried out heuristic homogenization of a single trilayer  piezoelectric element, based on its  scattering properties. Specifically, Eq.~(41f) therein suggests that the electromomentum coefficient depends on the contrast of $\Bb/\Aa$ between the layers, with some weight that depends on the rest of the properties. Accordingly, it appears that the optimal laminates maximize the contrast between the two layers which have the largest weight function.

The four optimal designs are similar to each other: the mapping from one optimal design to another is obtained by swapping the first and the third layer, and/or flipping the contrast of the mass density. 
This means that layers with minimal mass density are replaced with maximal mass density and vice versa; layers with maximal mass density are replaced with minimal mass density.
We refer to this transformation as the \emph{flipping} of $\rho$. 
Each of the two transformation steps alone flips the sign of $\effectiveprop{\couplingW}$. 

We demonstrate these two transformations using the laminates in Figure \ref{fig:Brho1} and in Figure \ref{fig:Brho4}, that share the same arrangement of the piezoelectric coefficient, but flipped values of the mass density, leading to the same magnitude with opposite sign of the electromomentum coefficient. The laminates in Figure \ref{fig:Brho1} and Figure \ref{fig:Brho2} have the mirrored arrangement of $\Bb$ and also flipped values of $\rho$, leading to the same value of $\effectiveprop{\couplingW}$. 

These observations correspond to the direction-dependent response of the medium \cite{rps20201wm}, captured by the anisotropic nature of $\effectiveprop{\couplingW}$. As noted by Pernas-Salomon et al. \cite{rps20201wm}, $\effectiveprop{\couplingW}$ flips sign when the coordinate system is reversed, which is equivalent to the inversion of the unit cell.

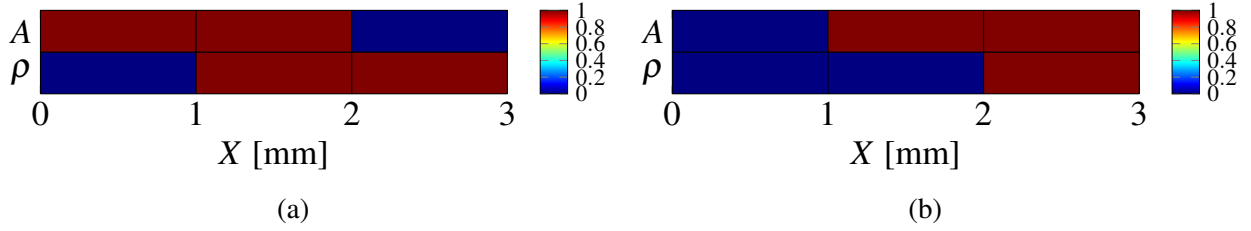
\begin{figure}[t!] 
\centering
\subfloat[]{\label{fig:Arho1} 
\resizebox{0.5\textwidth}{!}{
%
%
\begin{tikzpicture}

\begin{axis}[%
width=3.353in,
height=0.6in,
at={(0.562in,0.675in)},
scale only axis,
point meta min=0,
point meta max=1,
colormap/jet,
separate axis lines,
every outer x axis line/.append style={black},
every x tick label/.append style={font=\color{black}, font=\fontsize{18}{18}\selectfont},
every x tick/.append style={black},
xmin=0,
xmax=3,
xtick={0, 1, 2, 3},
xlabel={\fontsize{18}{18}\selectfont $X$ [mm]},
every outer y axis line/.append style={black},
every y tick label/.append style={font=\color{black}, font= \fontsize{18}{18}\selectfont},
every y tick/.append style={black},
ymin=0,
ymax=1,
ytick={0.25,0.75},
yticklabels={{$\rho$},{$\text{\it{} A}$}},
axis background/.style={fill=white},
colorbar
]

\addplot[area legend, table/row sep=crcr, patch, patch type=rectangle, shader=flat corner, draw=black, forget plot, patch table with point meta={%
0	1	2	3	0\\
4	5	6	7	1\\
8	9	10	11	1\\
12	13	14	15	1\\
16	17	18	19	1\\
20	21	22	23	0\\
}]
table[row sep=crcr] {%
x	y\\
0	0\\
1	0\\
1	0.5\\
0	0.5\\
0	0.5\\
1	0.5\\
1	1\\
0	1\\
1	0\\
2	0\\
2	0.5\\
1	0.5\\
1	0.5\\
2	0.5\\
2	1\\
1	1\\
2	0\\
3	0\\
3	0.5\\
2	0.5\\
2	0.5\\
3	0.5\\
3	1\\
2	1\\
};
\end{axis}
\end{tikzpicture}%
}}
\subfloat[]{\label{fig:Arho2} 
\resizebox{0.5\textwidth}{!}{
%
%
\begin{tikzpicture}

\begin{axis}[%
width=3.353in,
height=0.6in,
at={(0.562in,0.675in)},
scale only axis,
point meta min=0,
point meta max=1,
colormap/jet,
separate axis lines,
every outer x axis line/.append style={black},
every x tick label/.append style={font=\color{black}, font=\fontsize{18}{18}\selectfont},
every x tick/.append style={black},
xmin=0,
xmax=3,
xtick={0, 1, 2, 3},
xlabel={\fontsize{18}{18}\selectfont $X$ [mm]},
every outer y axis line/.append style={black},
every y tick label/.append style={font=\color{black}, font=\fontsize{18}{18}\selectfont},
every y tick/.append style={black},
ymin=0,
ymax=1,
ytick={0.25,0.75},
yticklabels={{$\rho$},{$\text{\it{} A}$}},
axis background/.style={fill=white},
colorbar
]

\addplot[area legend, table/row sep=crcr, patch, patch type=rectangle, shader=flat corner, draw=black, forget plot, patch table with point meta={%
0	1	2	3	0\\
4	5	6	7	0\\
8	9	10	11	0\\
12	13	14	15	1\\
16	17	18	19	1\\
20	21	22	23	1\\
}]
table[row sep=crcr] {%
x	y\\
0	0\\
1	0\\
1	0.5\\
0	0.5\\
0	0.5\\
1	0.5\\
1	1\\
0	1\\
1	0\\
2	0\\
2	0.5\\
1	0.5\\
1	0.5\\
2	0.5\\
2	1\\
1	1\\
2	0\\
3	0\\
3	0.5\\
2	0.5\\
2	0.5\\
3	0.5\\
3	1\\
2	1\\
};
\end{axis}
\end{tikzpicture}%
}}
\caption{Free material optimization over the dielectric and the mass density properties. Panels (a) and (b) show two equivalent optimal designs, where the properties were color-scaled as yellow (blue) color denotes the upper (lower) limit value, and the optimal $|\effectiveprop{\couplingW}|$ is $4.516\cdot10^{-6}$ Cs/m$^{3}$.}
\label{fig:Arho}
\end{figure}

Our  second case study optimizes over the mass density and dielectric coefficient. We find four  designs, two of which are displayed
in Figure \ref{fig:Arho}, which are related through the transformations mentioned earlier. The optimal coefficient is  $4.516\cdot10^{-6}$, which is smaller than the coefficient that was obtained when optimizing over $\rho$ and $B$, yet it is still three order of magnitude larger than that of the reference laminate. Interestingly, here all six properties attain  limiting values, not only five as in the optimization over $\rho$ and $B$, i.e., the dependency of $\tilde{W}$ on $A$ is different from its dependency on $B$.


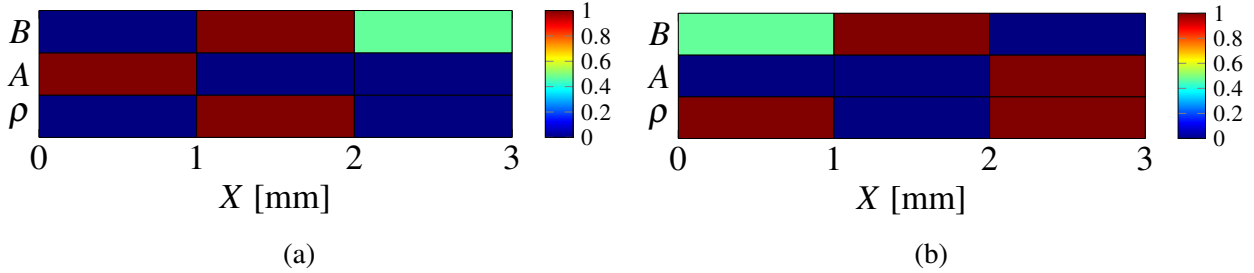
\begin{figure}[t] 
\subfloat[]{\label{fig:ABRHO1}
\resizebox{0.5\textwidth}{!}{
%
%
\begin{tikzpicture}

\begin{axis}[%
width=3.353in,
height=0.9in,
at={(0.562in,0.675in)},
scale only axis,
point meta min=0,
point meta max=1,
colormap/jet,
separate axis lines,
every outer x axis line/.append style={black},
every x tick label/.append style={font=\color{black}, font=\fontsize{18}{18}\selectfont},
every x tick/.append style={black},
xmin=0,
xmax=3,
xtick={0,1,2,3},
xlabel={\fontsize{18}{18}\selectfont $X$ [mm]},
every outer y axis line/.append style={black},
every y tick label/.append style={font=\color{black}, font=\fontsize{18}{18}\selectfont},
every y tick/.append style={black},
ymin=0,
ymax=1,
ytick={0.166666666666667,0.5,0.833333333333333},
yticklabels={{$\rho$},{$A$},{$B$}},
axis background/.style={fill=white},
colorbar
]

\addplot[area legend, table/row sep=crcr, patch, patch type=rectangle, shader=flat corner, draw=black, forget plot, patch table with point meta={%
0	1	2	3	0\\
4	5	6	7	1\\
8	9	10	11	0\\
12	13	14	15	1\\
16	17	18	19	0\\
20	21	22	23	1\\
24	25	26	27	0\\
28	29	30	31	0\\
32	33	34	35	0.4737\\
}]
table[row sep=crcr] {%
x	y\\
0	0\\
1	0\\
1	0.333333333333333\\
0	0.333333333333333\\
0	0.333333333333333\\
1	0.333333333333333\\
1	0.666666666666667\\
0	0.666666666666667\\
0	0.666666666666667\\
1	0.666666666666667\\
1	1\\
0	1\\
1	0\\
2	0\\
2	0.333333333333333\\
1	0.333333333333333\\
1	0.333333333333333\\
2	0.333333333333333\\
2	0.666666666666667\\
1	0.666666666666667\\
1	0.666666666666667\\
2	0.666666666666667\\
2	1\\
1	1\\
2	0\\
3	0\\
3	0.333333333333333\\
2	0.333333333333333\\
2	0.333333333333333\\
3	0.333333333333333\\
3	0.666666666666667\\
2	0.666666666666667\\
2	0.666666666666667\\
3	0.666666666666667\\
3	1\\
2	1\\
};
\end{axis}
\end{tikzpicture}%
}}
\subfloat[]{\label{fig:ABRHO2}
\resizebox{0.5\textwidth}{!}{
%
%
\begin{tikzpicture}

\begin{axis}[%
width=3.353in,
height=0.9in,
at={(0.562in,0.675in)},
scale only axis,
point meta min=0,
point meta max=1,
colormap/jet,
separate axis lines,
every outer x axis line/.append style={black},
every x tick label/.append style={font=\color{black}, font=\fontsize{18}{18}\selectfont},
every x tick/.append style={black},
xmin=0,
xmax=3,
xtick={0,1,2,3},
xlabel={\fontsize{18}{18}\selectfont $X$ [mm]},
every outer y axis line/.append style={black},
every y tick label/.append style={font=\color{black}, font=\fontsize{18}{18}\selectfont},
every y tick/.append style={black},
ymin=0,
ymax=1,
ytick={0.166666666666667,0.5,0.833333333333333},
yticklabels={{$\rho$},{$\text{\it{} A}$},{$\text{\it{} B}$}},
axis background/.style={fill=white},
colorbar
]

\addplot[area legend, table/row sep=crcr, patch, patch type=rectangle, shader=flat corner, draw=black, forget plot, patch table with point meta={%
0	1	2	3	1\\
4	5	6	7	0\\
8	9	10	11	0.4737\\
12	13	14	15	0\\
16	17	18	19	0\\
20	21	22	23	1\\
24	25	26	27	1\\
28	29	30	31	1\\
32	33	34	35	0\\
}]
table[row sep=crcr] {%
x	y\\
0	0\\
1	0\\
1	0.333333333333333\\
0	0.333333333333333\\
0	0.333333333333333\\
1	0.333333333333333\\
1	0.666666666666667\\
0	0.666666666666667\\
0	0.666666666666667\\
1	0.666666666666667\\
1	1\\
0	1\\
1	0\\
2	0\\
2	0.333333333333333\\
1	0.333333333333333\\
1	0.333333333333333\\
2	0.333333333333333\\
2	0.666666666666667\\
1	0.666666666666667\\
1	0.666666666666667\\
2	0.666666666666667\\
2	1\\
1	1\\
2	0\\
3	0\\
3	0.333333333333333\\
2	0.333333333333333\\
2	0.333333333333333\\
3	0.333333333333333\\
3	0.666666666666667\\
2	0.666666666666667\\
2	0.666666666666667\\
3	0.666666666666667\\
3	1\\
2	1\\
};
\end{axis}
\end{tikzpicture}%
}}
\caption{Free material optimization over the piezoelectric, the dielectric and the mass density properties. Panels (a) and (b) show two equivalent optimal designs, where the properties were color-scaled as yellow (blue) color denotes the upper (lower) limit value, and the optimal $|\effectiveprop{\couplingW}|$ is $1.0354\cdot10^{-5}$ Cs/m$^{3}$.}
\label{fig:ABRHO}
\end{figure}

The last optimization problem considers all these three material properties, i.e., $\Aa, \Bb$ and $\density$, as design variables. Figure \ref{fig:ABRHO} presents two optimal and equivalent designs of the design variables. Again, the mass density and dielectric coefficient of all the layers attain the limiting value, as well as the piezoelectric coefficient of two out of the three layers. The resultant electromomentum coefficient is larger by an additional order of magnitude, in comparison with the coefficient that was obtained when only pairs of material properties were optimized.

Finally, we report that additional computations (not shown here) with $\density$ or $\Bb/\Aa$ as prescribed constant values in the unit cell, lead to zero $\effectiveprop{\couplingW}$, regardless of the design variables chosen. The fact that constant $\Bb/\Aa$ yields zero $\tilde{W}$ is in agreement with the results of \citet{rps20201wm}. However, Eq.\ (41f) in the latter reference, which we recall is based on the homogenization of a single scatterer, implies that $\tilde{W}$ can be nonzero with constant $\density$, a result that is different from our current calculations, which are based on homogenization of periodic media.

\subsection{Complete free material optimization}
\label{sec:completeFMO}
 
\begin{figure}[t]
    \centering
    
    \begin{minipage}[b]{0.42\textwidth}
        \centering
        \resizebox{1.\textwidth}{!}{
%
%
\begin{tikzpicture}

\begin{axis}[%
width=3.353in,
height=1.2in,
at={(0.562in,0.675in)},
scale only axis,
point meta min=0,
point meta max=1,
colormap/jet,
separate axis lines,
every outer x axis line/.append style={black},
every x tick label/.append style={font=\color{black}, font=\fontsize{18}{18}\selectfont},
every x tick/.append style={black},
xmin=0,
xmax=3,
xtick={0, 1, 2, 3},
xlabel={\fontsize{18}{18}\selectfont $X$ [mm]},
every outer y axis line/.append style={black},
every y tick label/.append style={font=\color{black}, font=\fontsize{18}{18}\selectfont},
every y tick/.append style={black},
ymin=0,
ymax=1,
ytick={0.125,0.375,0.625,0.875},
yticklabels={{$\rho$},{$\text{\it{} C}$},{$\text{\it{} A}$},{$\text{\it{} B}$}},
axis background/.style={fill=white},
colorbar
]

\addplot[area legend, table/row sep=crcr, patch, patch type=rectangle, shader=flat corner, draw=black, forget plot, patch table with point meta={%
0	1	2	3	0\\
4	5	6	7	0\\
8	9	10	11	0\\
12	13	14	15	1\\
16	17	18	19	1\\
20	21	22	23	0\\
24	25	26	27	1\\
28	29	30	31	0\\
32	33	34	35	1\\
36	37	38	39	0\\
40	41	42	43	0\\
44	45	46	47	0.4416\\
}]
table[row sep=crcr] {%
x	y\\
0	0\\
1.404	0\\
1.404	0.25\\
0	0.25\\
0	0.25\\
1.404	0.25\\
1.404	0.5\\
0	0.5\\
0	0.5\\
1.404	0.5\\
1.404	0.75\\
0	0.75\\
0	0.75\\
1.404	0.75\\
1.404	1\\
0	1\\
1.404	0\\
1.7436	0\\
1.7436	0.25\\
1.404	0.25\\
1.404	0.25\\
1.7436	0.25\\
1.7436	0.5\\
1.404	0.5\\
1.404	0.5\\
1.7436	0.5\\
1.7436	0.75\\
1.404	0.75\\
1.404	0.75\\
1.7436	0.75\\
1.7436	1\\
1.404	1\\
1.7436	0\\
3	0\\
3	0.25\\
1.7436	0.25\\
1.7436	0.25\\
3	0.25\\
3	0.5\\
1.7436	0.5\\
1.7436	0.5\\
3	0.5\\
3	0.75\\
1.7436	0.75\\
1.7436	0.75\\
3	0.75\\
3	1\\
1.7436	1\\
};
\end{axis}
\end{tikzpicture}%
}
        \subcaption{}\label{fig:ABCRhoL}
        \resizebox{1.0\textwidth}{!}{
%
%
\begin{tikzpicture}

\begin{axis}[%
width=3.353in,
height=1.2in,
at={(0.562in,0.675in)},
scale only axis,
point meta min=0,
point meta max=1,
colormap/jet,
separate axis lines,
every outer x axis line/.append style={black},
every x tick label/.append style={font=\color{black}, font=\fontsize{18}{18}\selectfont},
every x tick/.append style={black},
xmin=0,
xmax=3,
xtick={0, 1, 2, 3},
xlabel={\fontsize{18}{18}\selectfont $X$ [mm]},
every outer y axis line/.append style={black},
every y tick label/.append style={font=\color{black}, font=\fontsize{18}{18}\selectfont},
every y tick/.append style={black},
ymin=0,
ymax=1,
ytick={0.125,0.375,0.625,0.875},
yticklabels={{$\rho$},{$\text{\it{} C}$},{$\text{\it{} A}$},{$\text{\it{} B}$}},
axis background/.style={fill=white},
colorbar
]

\addplot[area legend, table/row sep=crcr, patch, patch type=rectangle, shader=flat corner, draw=black, forget plot, patch table with point meta={%
0	1	2	3	0\\
4	5	6	7	0\\
8	9	10	11	0\\
12	13	14	15	0.4416\\
16	17	18	19	0\\
20	21	22	23	0\\
24	25	26	27	1\\
28	29	30	31	0\\
32	33	34	35	1\\
36	37	38	39	0\\
40	41	42	43	0\\
44	45	46	47	1\\
}]
table[row sep=crcr] {%
x	y\\
0	0\\
1.2564	0\\
1.2564	0.25\\
0	0.25\\
0	0.25\\
1.2564	0.25\\
1.2564	0.5\\
0	0.5\\
0	0.5\\
1.2564	0.5\\
1.2564	0.75\\
0	0.75\\
0	0.75\\
1.2564	0.75\\
1.2564	1\\
0	1\\
1.2564	0\\
1.596	0\\
1.596	0.25\\
1.2564	0.25\\
1.2564	0.25\\
1.596	0.25\\
1.596	0.5\\
1.2564	0.5\\
1.2564	0.5\\
1.596	0.5\\
1.596	0.75\\
1.2564	0.75\\
1.2564	0.75\\
1.596	0.75\\
1.596	1\\
1.2564	1\\
1.596	0\\
3	0\\
3	0.25\\
1.596	0.25\\
1.596	0.25\\
3	0.25\\
3	0.5\\
1.596	0.5\\
1.596	0.5\\
3	0.5\\
3	0.75\\
1.596	0.75\\
1.596	0.75\\
3	0.75\\
3	1\\
1.596	1\\
};
\end{axis}
\end{tikzpicture}%
}
        \subcaption{}\label{fig:ABCRhoL2}
    \end{minipage}
    \begin{minipage}[b]{.56\textwidth}
        \centering
        \resizebox{1.0\textwidth}{!}{\input{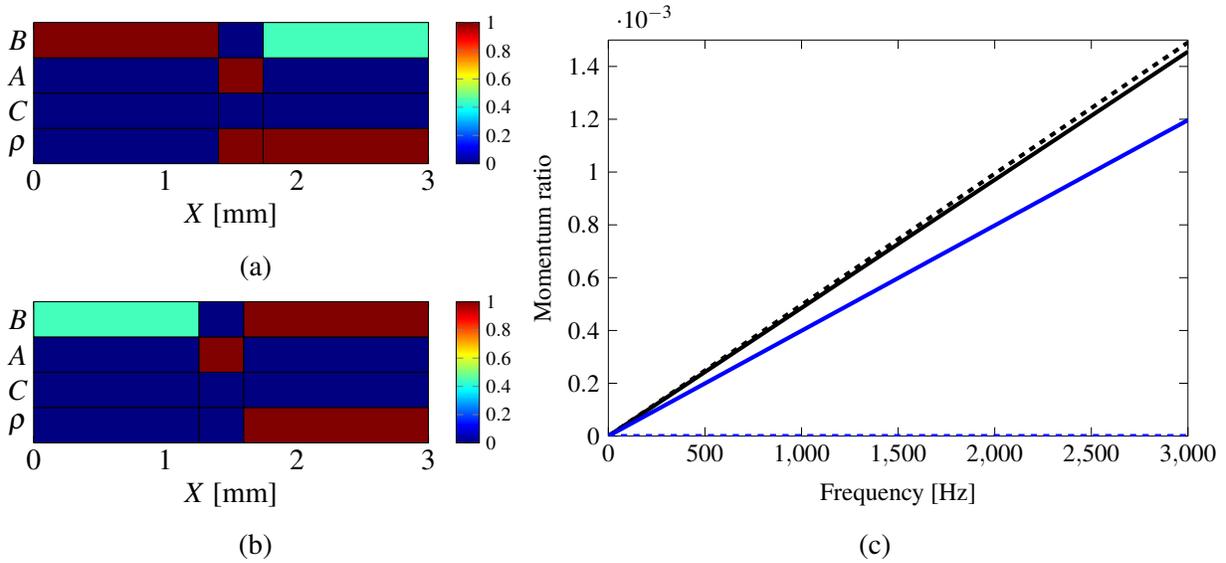}}
        \subcaption{}\label{fig:ratio}
    \end{minipage}
    
    \caption{Free material optimization over all the material and geometric properties. Panels (a) and (b) show two equivalent optimal designs, where the properties were color-scaled as yellow (blue) color denotes the upper (lower) limit value, and the optimal $|\effectiveprop{\couplingW}|$ is 5.7$\cdot10^{-5}$ Cs/m$^{3}$. Panel (c) displays the momentum ratios of the off-diagonal couplings: Solid blue (black) curve represents the Willis momentum ratio, namely,  $\frac{\effectiveprop{\couplingS}\effective{\derivative{\disp}}}{\effectiveprop{\rho}\effective{\derivativet{\disp}}}$ for the reference laminate (FMO optimal design), and dashed blue (black) curve denotes the electromomentum coupling momentum ratio, namely,  $\frac{\effectiveprop{\couplingW}\effective{\derivative{\elecpot}}}{\effectiveprop{\rho}\effective{\derivativet{\disp}}}$ for the reference laminate (FMO optimal design).}
    \label{fig:ABCRL}
\end{figure}

Figure \ref{fig:ABCRhoL}  presents the hypothetical optimal design that is obtained by removing all design restrictions. Accordingly, the design variables are the four material properties and the length of the layers. 
The resultant electromomentum coefficient is $5.7\cdot10^{-5}$ Cs/m$^{3}$, which is fourfold the coefficient that was obtained when the optimization was over the triplet $\rho,A$ and $B$.  All the material properties attain the limiting values, except the piezoelectric coefficient of one single layer.  Interestingly, the optimal stiffness is constant in the cell, the value of which is the lower limit, corresponding to a soft laminate. The optimal lengths yield a thin layer (volume fraction of $11\%$),  in-between two thick layers (volume fractions of $47\%$ and $42\%$). This result resembles the one that we obtained in the length optimization problem. Note that, again, equivalent designs can be obtained by mirroring the periodic cell and/or flipping the mass density  values. Such equivalent design is shown in Figure \ref{fig:ABCRhoL2}.


Interestingly,  the optimization process not only increases $\effectiveprop{\couplingW}$, but also increases the relative contribution of both the electromomentum and the Willis effects to the effective linear momentum. To show this, we compare in Figure \ref{fig:ratio} the momentum ratios  $\frac{\effectiveprop{\couplingS}\effective{\derivative{\disp}}}{\effectiveprop{\rho}\effective{\derivativet{\disp}}}$ (solid lines) and $\frac{\effectiveprop{\couplingW}\effective{\derivative{\elecpot}}}{\effectiveprop{\rho}\effective{\derivativet{\disp}}}$ (dashed lines), in the reference laminate (blue lines)  and the FMO design-based laminate (black lines), as functions of the frequency. (The reader is referred to Appendix C in the Ref. \cite{PernasSalomon2019JMPS} for details in the calculations of $\effective{\derivativet{\disp}}$ and $\effective{\derivative{\elecpot}}$.) We also observe that while $\effectiveprop{\couplingW}\effective{\derivative{\elecpot}}$ in the reference laminate is negligible with respect to $\effectiveprop{\couplingS}\effective{\derivativet{\disp}}$, it surpasses $\effectiveprop{\couplingS}\effective{\derivativet{\disp}}$ in the optimized laminate. 

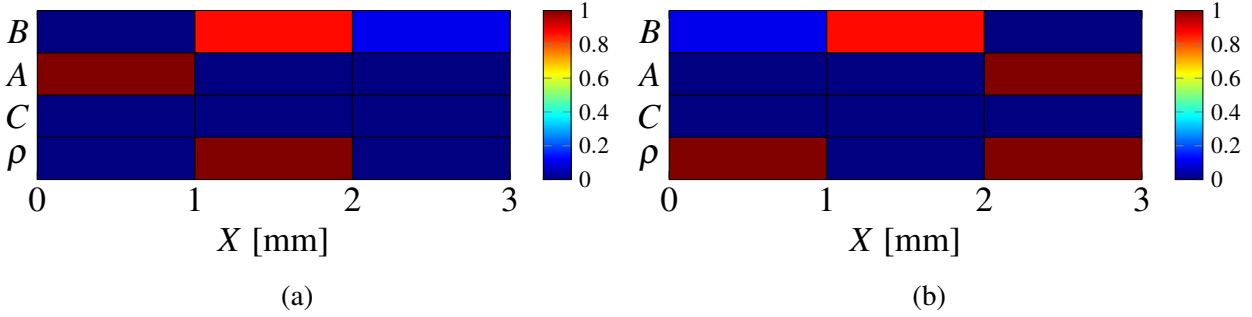
\begin{figure}[t] 
\subfloat[]{\label{fig:ABCRho1}
\resizebox{0.5\textwidth}{!}{
%
%
\begin{tikzpicture}

\begin{axis}[%
width=3.353in,
height=1.2in,
at={(0.562in,0.675in)},
scale only axis,
point meta min=0,
point meta max=1,
colormap/jet,
separate axis lines,
every outer x axis line/.append style={black},
every x tick label/.append style={font=\color{black}, font=\fontsize{18}{18}\selectfont},
every x tick/.append style={black},
xmin=0,
xmax=3,
xtick={0, 1, 2, 3},
xlabel={\fontsize{18}{18}\selectfont $X$ [mm]},
every outer y axis line/.append style={black},
every y tick label/.append style={font=\color{black}, font=\fontsize{18}{18}\selectfont},
every y tick/.append style={black},
ymin=0,
ymax=1,
ytick={0.125,0.375,0.625,0.875},
yticklabels={{$\rho$},{$C$},{$A$},{$B$}},
axis background/.style={fill=white},
colorbar
]

\addplot[area legend, table/row sep=crcr, patch, patch type=rectangle, shader=flat corner, draw=black, forget plot, patch table with point meta={%
0	1	2	3	0\\
4	5	6	7	0\\
8	9	10	11	1\\
12	13	14	15	0\\
16	17	18	19	1\\
20	21	22	23	0\\
24	25	26	27	0\\
28	29	30	31	0.8668\\
32	33	34	35	0\\
36	37	38	39	0\\
40	41	42	43	0\\
44	45	46	47	0.1061\\
}]
table[row sep=crcr] {%
x	y\\
0	0\\
1	0\\
1	0.25\\
0	0.25\\
0	0.25\\
1	0.25\\
1	0.5\\
0	0.5\\
0	0.5\\
1	0.5\\
1	0.75\\
0	0.75\\
0	0.75\\
1	0.75\\
1	1\\
0	1\\
1	0\\
2	0\\
2	0.25\\
1	0.25\\
1	0.25\\
2	0.25\\
2	0.5\\
1	0.5\\
1	0.5\\
2	0.5\\
2	0.75\\
1	0.75\\
1	0.75\\
2	0.75\\
2	1\\
1	1\\
2	0\\
3	0\\
3	0.25\\
2	0.25\\
2	0.25\\
3	0.25\\
3	0.5\\
2	0.5\\
2	0.5\\
3	0.5\\
3	0.75\\
2	0.75\\
2	0.75\\
3	0.75\\
3	1\\
2	1\\
};
\end{axis}
\end{tikzpicture}%
}}
\subfloat[]{\label{fig:ABCRho2}
\resizebox{0.5\textwidth}{!}{
%
%
\begin{tikzpicture}

\begin{axis}[%
width=3.353in,
height=1.2in,
at={(0.562in,0.675in)},
scale only axis,
point meta min=0,
point meta max=1,
colormap/jet,
separate axis lines,
every outer x axis line/.append style={black},
every x tick label/.append style={font=\color{black}, font=\fontsize{18}{18}\selectfont},
every x tick/.append style={black},
xmin=0,
xmax=3,
xtick={0, 1, 2, 3},
xlabel={\fontsize{18}{18}\selectfont $X$ [mm]},
every outer y axis line/.append style={black},
every y tick label/.append style={font=\color{black}, font=\fontsize{18}{18}\selectfont},
every y tick/.append style={black},
ymin=0,
ymax=1,
ytick={0.125,0.375,0.625,0.875},
yticklabels={{$\rho$},{$C$},{$A$},{$B$}},
axis background/.style={fill=white},
colorbar
]

\addplot[area legend, table/row sep=crcr, patch, patch type=rectangle, shader=flat corner, draw=black, forget plot, patch table with point meta={%
0	1	2	3	1\\
4	5	6	7	0\\
8	9	10	11	0\\
12	13	14	15	0.1061\\
16	17	18	19	0\\
20	21	22	23	0\\
24	25	26	27	0\\
28	29	30	31	0.8668\\
32	33	34	35	1\\
36	37	38	39	0\\
40	41	42	43	1\\
44	45	46	47	0\\
}]
table[row sep=crcr] {%
x	y\\
0	0\\
1	0\\
1	0.25\\
0	0.25\\
0	0.25\\
1	0.25\\
1	0.5\\
0	0.5\\
0	0.5\\
1	0.5\\
1	0.75\\
0	0.75\\
0	0.75\\
1	0.75\\
1	1\\
0	1\\
1	0\\
2	0\\
2	0.25\\
1	0.25\\
1	0.25\\
2	0.25\\
2	0.5\\
1	0.5\\
1	0.5\\
2	0.5\\
2	0.75\\
1	0.75\\
1	0.75\\
2	0.75\\
2	1\\
1	1\\
2	0\\
3	0\\
3	0.25\\
2	0.25\\
2	0.25\\
3	0.25\\
3	0.5\\
2	0.5\\
2	0.5\\
3	0.5\\
3	0.75\\
2	0.75\\
2	0.75\\
3	0.75\\
3	1\\
2	1\\
};
\end{axis}
\end{tikzpicture}%
}}
\caption{Free material optimization over all the material properties with the length held fixed. Panels (a) and (b) show two equivalent optimal designs, where the properties were color-scaled as yellow (blue) color denotes the upper (lower) limit value, and the optimal $|\effectiveprop{\couplingW}|$ is 4.192$\cdot10^{-5}$ Cs/m$^{3}$.}
\label{fig:ABCRHO}
\end{figure}

To quantify the contribution of the geometrical variables to the optimization, we optimize next only over the four material properties, keeping the length of each layer fixed and equal to 1$\,$mm. The resultant electromomentum coefficient is $4.192\cdot10^{-5}$ Cs/m$^3$. This value is about 71$\%$ of the coefficient that was obtained when the geometrical and material parameters were included in the design space. Two of the equivalent, optimal designs are displayed in Figure \ref{fig:ABCRHO}. These laminates are different from those that were obtained so far, as none of the layers exhibits a maximal piezoelectric coefficient.

\subsection{Optimization over the designable mass density and stiffness }
\label{sec:BestApp}
\begin{figure}[t]
\centering
%
%
\begin{tikzpicture}

\begin{axis}[%
width=3.353in,
height=0.6in,
at={(0.562in,0.675in)},
scale only axis,
point meta min=0,
point meta max=1,
colormap/jet,
separate axis lines,
every outer x axis line/.append style={black},
every x tick label/.append style={font=\color{black}},
every x tick/.append style={black},
xmin=0,
xmax=3,
xtick={0, 1, 2, 3},
xlabel={X [mm]},
every outer y axis line/.append style={black},
every y tick label/.append style={font=\color{black}},
every y tick/.append style={black},
ymin=0,
ymax=1,
ytick={0.25,0.75},
yticklabels={{$\rho$},{$\text{\it{} C}$}},
axis background/.style={fill=white},
colorbar
]

\addplot[area legend, table/row sep=crcr, patch, patch type=rectangle, shader=flat corner, draw=black, forget plot, patch table with point meta={%
0	1	2	3	1\\
4	5	6	7	0\\
8	9	10	11	0\\
12	13	14	15	0\\
16	17	18	19	0\\
20	21	22	23	0\\
}]
table[row sep=crcr] {%
x	y\\
0	0\\
2.0489	0\\
2.0489	0.5\\
0	0.5\\
0	0.5\\
2.0489	0.5\\
2.0489	1\\
0	1\\
2.0489	0\\
2.1762	0\\
2.1762	0.5\\
2.0489	0.5\\
2.0489	0.5\\
2.1762	0.5\\
2.1762	1\\
2.0489	1\\
2.1762	0\\
3	0\\
3	0.5\\
2.1762	0.5\\
2.1762	0.5\\
3	0.5\\
3	1\\
2.1762	1\\
};
\end{axis}
\end{tikzpicture}%
\caption{Optimization over the designable mass density and stiffness. The optimal design is the result of optimizing the composition of the DMO optimal result (Figure \ref{fig:DMO}) over continuous mass density, stiffness and length. The lengths of the laminates are $2.0489$ mm, $0.1273$ mm and $0.8238$ mm. The properties were color-scaled as yellow (blue) color denotes the upper (lower) limit value. The optimal objective is $1.282\cdot10^{-6}$ Cs/m$^3$.}
\label{fig:GAresultOpt}
\end{figure}
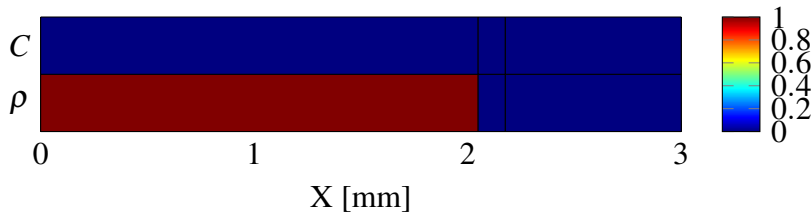

The results of the previous section  provide a hypothetical upper bound for the electromomentum coefficient, which is accessible only if all material properties can be engineered independently. The significant progress on the design of engineered stiffness and mass density \cite{WU2021,Dalela2021},  motivates us to term the optimal design of  $\stiffness$ and $\density$ as the applicable design. Figure \ref{fig:GAresultOpt} shows the solution of this optimization over  $\stiffness$ and $\density$, when the rest  of the  properties are chosen according to the DMO optimal result (Figure \ref{fig:DMO}). The resultant electromomentum coefficient is   $1.282\cdot10^{-6}$ Cs/m$^3$, which is only one order of magnitude lower than the hypothetical upper bound. Here again, the optimization process delivers a constant minimal stiffness, and maximizes the contrast of $\density$ in the cell, as it attains the lowest limit in layers two and three, and the upper limit in layer one.

A summary of the optimal electromomentum coefficients, together with the reference laminate and the improvement relatively to it, is given in Table \ref{tab:conclude2}.
 
\begin{table}[t]
\begin{center}
\begin{tabular}{l|ccc}
\hline
 &  $|\effectiveprop{W}|$ [Cs/m$^3$] & Relative improvement ratio\\
\hline
The reference laminate &  $2.342\cdot10^{-9}$ & -  \\
DMO &  $6.312\cdot10^{-7}$ &  $269.51$\\
Length optimization  &  $9.52\cdot10^{-9}$ & $ 4.06 $\\
FMO: B and $\rho$ &  $5.687\cdot10^{-6}$ & $2428.26$\\
FMO: A and $ \rho$ &  $4.516\cdot10^{-6}$ & $1928.27$\\
FMO: B, A and $ \rho$ &   $1.0354\cdot10^{-5}$ & $4421.01$\\
FMO: all properties &  $4.192\cdot10^{-5}$ & $17899.23$\\
FMO: all properties and length &  $5.7\cdot10^{-5}$ & $24338.17$\\
Synthesizing C and $\rho$ &  $1.282\cdot10^{-6}$ & $547.40$\\
\hline\hline
\end{tabular}
\caption{The electromomentum coefficient of the reference laminate (equispaced Al$_{2}$O$_{3}$, BaTiO$_{3}$ and PMMA layers);  the optimal laminates that were reported in this work, and the relative improvement with respect to the reference laminate coefficient.}
\label{tab:conclude2} 
\end{center}
\end{table}

\section{Summary}
\label{sec:sum}

A prominent challenge in engineering today is to  control elastic waves; metamaterials exhibiting the  electromomentum effect \cite{PernasSalomon2019JMPS} offer a way to tune the phase velocity and generate asymmetry in the phase angle of such waves \cite{PernasSalomon2019JMPS,rps20201wm}. To make these phenomena more pronounced, it is required to design these metamaterials with large electromomentum coefficient. In this work, we have utilized discrete material-, topology- and free material optimization methods to maximize the electromomentum coupling of periodic piezoelectric trilayer laminates in the low frequency, long-wavelength limit. These optimization methods are essentially inverse homogenization methods, since we have formulated the objective function, namely, the electromomentum coefficient, as the end result of a rigorous homogenization process \cite{PernasSalomon2019JMPS,muhafra2021}. 

The discrete material optimization, which uses a predefined set of real materials as its design space, has yielded an electromomentum coefficient that is two orders of magnitude larger than the arbitrary laminate that was first analyzed by \citet{PernasSalomon2019JMPS}. We observed that the optimization yields a choice of materials that tends to maximize the contrast between their electromechanical properties, in a unit cell that comprises one thin layer in-between two thick layers. 

The free material optimization uses a design space with continuous variables, the values of which are bounded in-between the extreme values of the materials considered earlier. We have employed gradient-based algorithms that were implemented together with a sensitivity analysis of the objective function. Using these algorithms, we have solved  optimization problems whose design variables are pairs of the mass density and one of the electromechanical properties: the dielectric-, piezoelectric- or stiffness coefficients. The rest of the properties were set to the properties of the reference laminate. We have found that the optimal values of each one of these pairs enlarge the electromomentum coefficient by another order of magnitude, in comparison with the optimal laminate that the discrete material optimization delivered. We observed that the pair that demonstrated the best improvement was with the piezoelectric coefficient. We further found that when all material properties are designable, the electromomentum coefficient is enlarged by another order of magnitude. 

The optimal laminates reveal how the electromomentum  depends on the contrast in the electromechanical properties between the layers. Specifically, we observed that the optimization tends to maximize the electromechanical properties in certain layers, and minimize the same properties in the remaining layers. Two exceptions are the piezoelectric coefficient, which, while tending to the two extreme values in two layers, tends to an intermediate value at the remaining layer; and the stiffness, which the optimization  minimizes throughout the whole cell. The optimal laminates also show that the geometry that maximizes the electromomentum coefficient is of a thin layer, in-between two thick layers.

Collectively, our results provide guidelines for the design of metamaterials with  maximum electromomentum coefficient, paving the way for future work on their integration in wave control applications.

\section*{Acknowledgments}

The authors wish to thank Alon Landmann for providing the controls over convergence of GA in Eq.~\eqref{eq:Conf}. This research was supported by the Ministry of Science and Technology, Israel (Grant no. 880011), and by the Israel Science Foundation, funded by the Israel Academy of Sciences and Humanities (Grant no. 2061/20).

\appendixpage
\appendix
\section{The convergence of the PWE method}
\label{appendix2}

\begin{figure}[t]
\centering
\includegraphics[width=0.6\textwidth]{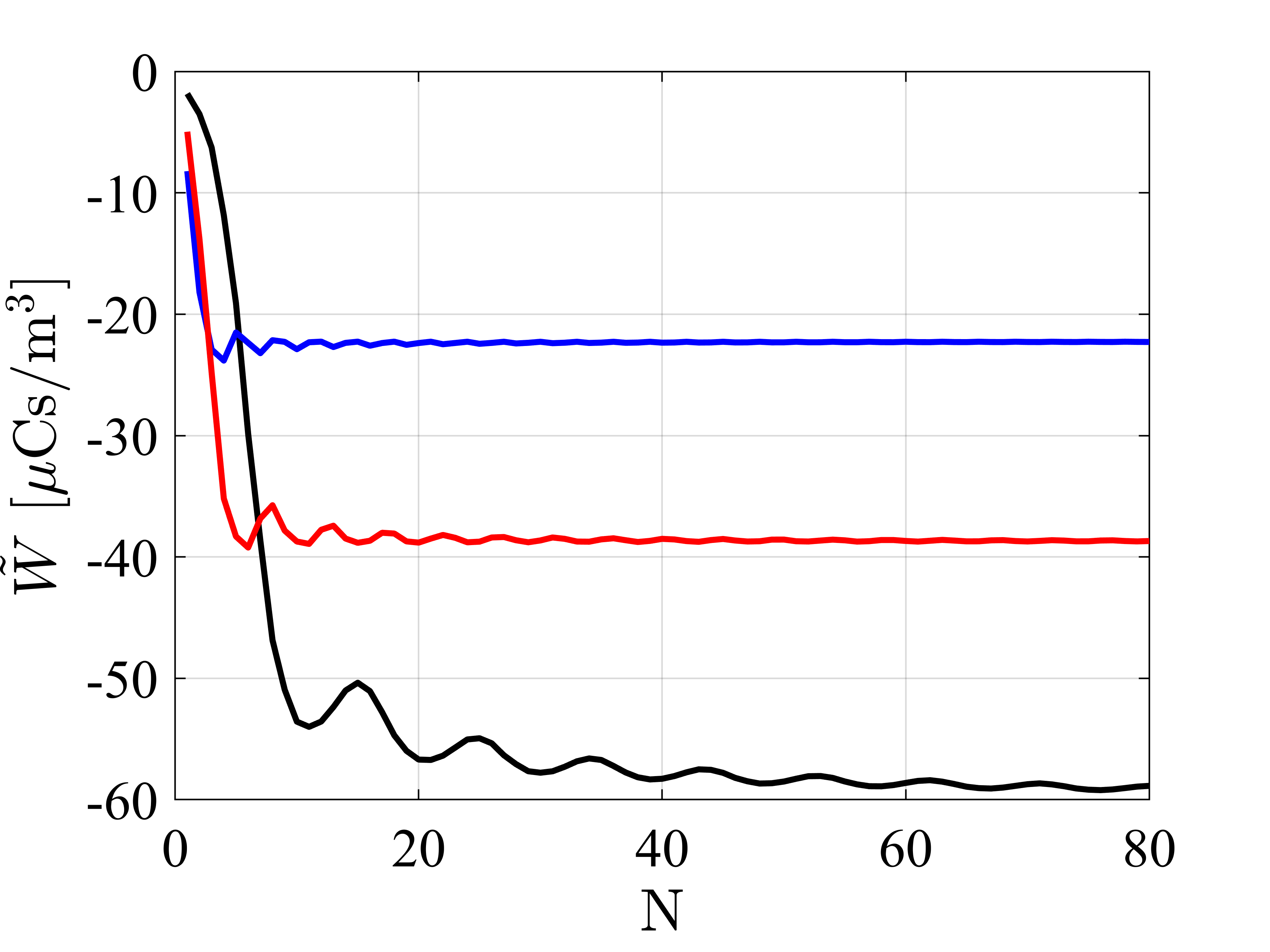}
\caption{The electromomentum coupling versus the number of Fourier waves used. The black, red and blue curves denote the high, medium and low contrast structures, respectively. The three curves show convergence at N = $35$.}
\label{fig:PWE_Convergence}
\end{figure}

To examine the convergence of the method, the
electromomentum coupling was evaluated for a combination of three fixed materials. Results show convergence at $30$ Fourier waves,  as shown in Figure \ref{fig:PWE_Convergence}.  Moreover, to examine the rate of convergence, three different divisions were considered; one is equally divided, one with one short layer compared to the other two layers and the third is the average of the previous two. As expected, the higher the contrast between the length of each layer, the slower the rate of convergence of the method. 

\section{Detailed expressions for the matrices in the PWE method}
\label{appendix1}

The matrices can be divided into sub-matrices following
\begin{equation}
\begin{aligned} & \assembly{\prop}=\left(\begin{array}{ccc}
\ij{\prop}11 & \ij{\prop}12 & \ij{\prop}13\\
\ij{\prop}21 & \ij{\prop}22 & \ij{\prop}23\\
\ij{\prop}31 & \ij{\prop}32 & \ij{\prop}33
\end{array}\right),\assembly{\opgov}=\left(\begin{array}{cc}
\ij{\opgov}11 & \ij{\opgov}12\\
\ij{\opgov}21 & \ij{\opgov}22\\
\ij{\opgov}31 & \ij{\opgov}32
\end{array}\right)^{\mathsf{T}},\assembly{\kineticv}=\left(\begin{array}{c}
\ijv{\kineticv}1\\
\ijv{\kineticv}2\\
\ijv{\kineticv}3
\end{array}\right),\\
 & \assembly{\opcon}=\left(\begin{array}{cc}
\ij{\opcon}11 & \ij{\opcon}12\\
\ij{\opcon}21 & \ij{\opcon}22\\
\ij{\opcon}31 & \ij{\opcon}32
\end{array}\right),\assembly{\kinematicv}=\left(\begin{array}{c}
\ijv{\kinematicv}1\\
\ijv{\kinematicv}2
\end{array}\right),\assembly{\strainsov}=\left(\begin{array}{c}
\ijv{\strainsov}1\\
\ijv{\strainsov}2\\
\ijv{\strainsov}3
\end{array}\right),\assembly{\forces}=\left(\begin{array}{c}
\ijv{\forces}1\\
\ijv{\forces}2
\end{array}\right),
\end{aligned}
\end{equation}
where the components of the matrix $\assembly{\prop}$ are simply a sub-matrices of the material properties, namely
\begin{equation}
\begin{aligned}\ijggt{\prop}11=\coeffggt{\stiffness}, & \ijggt{\prop}12=\coeffggt{\Bb},\\
\ijggt{\prop}21=\coeffggt{\Bb}, & \ijggt{\prop}22=-\coeffggt{\Aa},\\
\ijggt{\prop}33=\coeffggt{\density}, & \ijggt{\prop}13=\ijggt{\prop}23=\ijggt{\prop}31=\ijggt{\prop}32=\coeffggt 0,
\end{aligned}
\end{equation}
and as has been mentioned before,  $\assembly{\opgov}$ is composed of three diagonal matrices and three zero matrices, namely
\begin{equation}
\begin{aligned} & \ijggt{\opgov}11=\ijggt{\opgov}22=\coeffggt{\delta}i\kG,\\
 & \ijggt{\opgov}31=\coeffggt{\delta}i\angvel,\\
 & \ijggt{\opgov}12=\ijggt{\opgov}21=\ijggt{\opgov}32=\coeffggt 0,
\end{aligned}
\end{equation}
in the same manner, $\assembly{\opcon}$ is in the form
\begin{equation}
\begin{aligned} & \ijggt{\opcon}11=\ijggt{\opcon}22=\coeffggt{\delta}i\kGt,\\
 & \ijggt{\opcon}31=-\coeffggt{\delta}i\angvel,\\
 & \ijggt{\opcon}12=\ijggt{\opcon}21=\ijggt{\opcon}32=\coeffggt 0.
\end{aligned}
\end{equation}
The description of the vectors is similar, where $\assembly{\kineticv}$ is
\begin{equation}
\begin{aligned} & \ijvg{\kineticv}1=\coeffg{\stress},
 & \ijvg{\kineticv}2=\coeffg{\elecdisp},
 & \ijvg{\kineticv}3=\coeffg{\momentum},
\end{aligned}
\end{equation}
and $\assembly{\kinematicv}$
\begin{equation}
\begin{aligned} & \ijvgt{\kinematicv}1=\coeffgt{\disp},
 & \ijvgt{\kinematicv}2=\coeffgt{\elecpot},
\end{aligned}
\end{equation}
and $\assembly{\forces}$ definitely
\begin{equation}
\begin{aligned} & \ijvg{\forces}1=\coeffg{\force},
 & \ijvg{\forces}2=-\coeffg{\charge},
\end{aligned}
\end{equation}
and for $\assembly{\strainsov}$

\begin{equation}
\begin{aligned} &
\ijv{\strainsov}1=\coeffg{\strainso}, &
\ijv{\strainsov}2=0, &
\ijv{\strainsov}3=0.
\end{aligned}
\end{equation}

\pagebreak
\bibliography{sample.bib,bibtexfiletot.bib}
\bibliographystyle{unsrtnat}

\end{document}